\documentclass[a4paper,12pt,twoside]{article}
\usepackage{amsmath,amssymb,graphicx,psfrag}
\usepackage{layout}
\usepackage{color}
\usepackage{a4wide}                     
\usepackage{graphicx}
\usepackage{rotating}
\usepackage{cite}

\def\ra{\rightarrow} 
\newcommand{\beq}{\begin{equation}}
\newcommand{\eeq}{\end{equation}}
\def\gs{\mathrel{ \rlap{\raise
0.511ex \hbox{$>$}}{\lower 0.511ex \hbox{$\sim$}}}} \def\ls{\mathrel{
\rlap{\raise 0.511ex \hbox{$<$}}{\lower 0.511ex \hbox{$\sim$}}}}

\newcommand{\ba}{\begin{array}{c}}
\newcommand{\baz}{\begin{array}{cc}}
\newcommand{\bad}{\begin{array}{ccc}}
\newcommand{\bea}{\begin{equation} \begin{array}{c}}
\newcommand{\eea}{ \end{array} \end{equation}}
\newcommand{\ea}{\end{array}} 

\newcommand{\dmsol}{\mbox{$\Delta m^2_{\odot}$}}
\newcommand{\dma}{\mbox{$\Delta m^2_{\rm A}$}}

 \def\ra{\rightarrow}

\def\ra{\rightarrow} 
\def\gtap{\mathrel{ \rlap{\raise 0.511ex \hbox{$>$}}{\lower 0.511ex
   \hbox{$\sim$}}}} 
\def\ltap{\mathrel{ \rlap{\raise 0.511ex
   \hbox{$<$}}{\lower 0.511ex \hbox{$\sim$}}}}
   \newcommand{\deltaatm}{\mbox{$\Delta m^2_{31}$}}
   \newcommand{\deltasol}{\mbox{$ \Delta m^2_{21}$}}
   
   \newcommand{\betabeta}{\mbox{$(\beta \beta)_{0 \nu}$}}

   \newcommand{\hbeta}{$\mbox{}^3 {\rm H}$ $\beta$-decay }

\newcommand{\pmns}{\mbox{$ U_{\rm PMNS}$}}

\def\ie{\hbox{\it i.e.}{}}
\def\eg{\hbox{\it e.g.}{}}
\def\etc{\hbox{\it etc}{}}
\renewcommand{\thefootnote}{\fnsymbol{footnote}}
\begin{document}
\begin{titlepage}
\hfill
\vbox{
    \halign{#\hfil        \cr
           SISSA 59/2005/EP \cr
           IC/2005/045    \cr
           hep-ph/0508243\cr}}
\vspace*{10mm}
\begin{center}
{\large{\bf Majorana CP-Violating Phases, 
RG Running of Neutrino Mixing Parameters 
and Charged Lepton Flavour Violating Decays 
}}
\vspace*{7mm}
{\ S.~T.~Petcov ${}^{a, b)}$}\footnote[1]{Also at: Institute
of Nuclear Research and Nuclear Energy, Bulgarian Academy %
of Sciences, 1784 Sofia, %
Bulgaria}
{\ T.~Shindou ${}^{a, b)}$}\footnote[2]{E-mail: shindou@he.sissa.it},
{\ Y.~Takanishi ${}^{c)}$}\footnote[3]{E-mail: yasutaka@ictp.trieste.it},

\vspace*{.2cm}

{${}^{a)}${\it Scuola Internazionale Superiore di Studi
Avanzati, I-34014 Trieste, Italy}
\vskip .15cm
${}^{b)}${\it Istituto Nazionale di Fisica Nucleare,
Sezione di Trieste, I-34014 Trieste, Italy}
\vskip .15cm
${}^{c)}${\it The Abdus Salam International Centre for Theoretical
Physics, Strada Costiera 11, I-34100 Trieste, Italy}}
\end{center}
\begin{abstract}
  We consider the MSSM with see-saw mechanism of neutrino mass
  generation and soft SUSY breaking with flavour-universal 
  boundary conditions at the GUT scale, 
  in which the lepton flavour violating (LFV) decays $\mu
  \rightarrow e + \gamma$, $\tau \rightarrow \mu + \gamma$, $\etc$.,
  are predicted with rates that can be within the reach of present and
  planned experiments.  These predictions depend critically on the
  matrix of neutrino Yukawa couplings $\mathbf{Y_{\nu}}$ which can be
  expressed in terms of the light and heavy right-handed (RH) neutrino
  masses, neutrino mixing matrix $\pmns$, and an orthogonal matrix
  $\mathbf{R}$.  We investigate the effects of Majorana CP-violation
  phases in $\pmns$, and of the RG running of light neutrino masses
  and mixing angles from $M_Z$ to the RH Majorana neutrino mass scale
  $M_R$, on the predictions for the rates of LFV decays $\mu
  \rightarrow e + \gamma$, $\tau \rightarrow \mu + \gamma$ and $\tau
  \rightarrow e + \gamma$.  Results for neutrino mass spectrum with
  normal hierarchy, values of the lightest $\nu$-mass in the range $0
  \leq m_1 \leq 0.30$ eV, and quasi-degenerate heavy RH Majorana
  neutrinos in the cases of $\mathbf{R} = \mathbf{1}$ and complex
  matrix $\mathbf{R}$ are presented.  We find that the effects of the
  Majorana CP-violation phases and of the RG evolution of neutrino
  mixing parameters can change by few orders of magnitude the
  predicted rates of the LFV decays $\mu \to e + \gamma$ and $\tau \to
  e + \gamma$.  The impact of these effects on the $\tau \to \mu +
  \gamma$ decay rate is typically smaller and only possible for $m_1
  \gtap 0.10$ eV.  If the RG running effects are negligible, 
   in a large region of soft SUSY breaking
   parameter space the ratio of the branching ratios of the $\mu \to e
   + \gamma$ and $\tau \to e + \gamma$ ($\tau \to \mu + \gamma$) decays
   is entirely determined in the case of $\mathbf{R} \cong \mathbf{1}$
   by the values of the neutrino mixing parameters at $M_Z$.
  
\end{abstract}
\vskip 0.5cm

\today
\end{titlepage}

\newpage
\renewcommand{\thefootnote}{\arabic{footnote}}
\setcounter{footnote}{0}
\setcounter{page}{1}

\section{{\large Introduction}}
\hskip 1.0truecm 
The experiments with solar, 
atmospheric, reactor and accelerator 
neutrinos~\cite{sol,SKsolaratm,SNO123,KamLAND,K2K}
have provided during the last several years 
compelling evidence for the existence 
of neutrino oscillations caused by nonzero neutrino 
masses and neutrino mixing. The $\nu$-oscillation
data obtained in the solar neutrino, 
atmospheric neutrino,
KamLAND (KL) and K2K experiments 
imply the existence of 3-$\nu$ mixing
in the weak charged-lepton current 
(see, $\eg$,~\cite{STPNu04}):
\begin{equation}
\nu_{l \mathrm{L}}  = \sum_{j=1}^{3} U_{l j} \, \nu_{j \mathrm{L}},~~
l  = e,\mu,\tau,
\label{3numix}
\end{equation}
%
\noindent where 
$\nu_{lL}$ are the flavour neutrino fields, $\nu_{j \mathrm{L}}$ is
the field of neutrino $\nu_j$ having a mass $m_j$ and $U$ is the
Pontecorvo--Maki--Nakagawa--Sakata (PMNS) mixing 
matrix~\cite{BPont57}, $U \equiv \pmns$.  It follows from the results of the
neutrino oscillation and \hbeta experiments~\cite{MoscowH3Mainz} that
the neutrino masses $m_j$ are much smaller than the masses of the
charged leptons and quarks: $m_{j} < 2.3$ eV 
(95\% C.L.)~\footnote{More stringent upper limit on $m_j$ follows from the
  constraints on the sum of neutrino masses obtained from
  cosmological/astrophysical observations, namely, the CMB data of the
  WMAP experiment combined with data from large scale structure
  surveys (2dFGRS, SDSS)~\cite{WMAPnu}: $\sum_{j} m_j < (0.7 - 2.0)$
  eV (95\% C.L.), where we have included a conservative estimate of
  the uncertainty in the upper limit (see, $\eg$,~\cite{Hanne03}).  }.

A natural explanation of the smallness of neutrino masses is provided
by the see-saw mechanism of neutrino mass generation~\cite{seesaw}.
The see-saw mechanism predicts the light massive neutrinos $\nu_j$ to
be Majorana particles. An integral part of the mechanism are the heavy
right-handed (RH) Majorana neutrinos. Being singlets with respect to
the Standard Theory gauge symmetry group~\cite{Pont67}, the RH
neutrinos can acquire Majorana masses which are much larger than the
electroweak symmetry breaking scale. The CP-violating decays of the
heavy RH Majorana neutrinos in the Early Universe could generate,
through the leptogenesis scenario, the observed baryon asymmetry of
the Universe~\cite{LeptoG}.

In grand unified theories (GUT) the masses of the heavy RH Majorana
neutrinos are typically by a few to several orders of magnitude
smaller than the scale of unification of the electroweak and strong
interactions, $M_{\rm GUT} \sim 2\times 10^{16}$ GeV.  The presence of the
electroweak symmetry breaking scale and of the scale of RH neutrino
masses $M_R$ in a theory~\footnote{In the case of three heavy RH
  Majorana neutrinos $N_j$ with masses $M_j$ we will consider, $M_R$
  is determined by the mass of the lightest RH Majorana neutrino, $M_R
  = {\rm min}(M_j)$.}  can lead to a hierarchy problem.  In SUSY (GUT)
theories this problem could be avoided: the hierarchy between the two mass
(energy) scales is stabilised. Thus, the SUSY (GUT) theories with
see-saw mechanism provide an internally consistent framework for
generation of light neutrino masses $m_j$ and of the baryon asymmetry
of the Universe.
  
The existence of the flavour neutrino mixing, eq.~(\ref{3numix}),
implies that the individual lepton charges, $L_l$, $l =e,\mu,\tau$,
are not conserved (see, $\eg$,~\cite{BiPet87}).  Thus, lepton charge
non-conserving processes like $\mu^- \rightarrow e^- + \gamma$,
$\mu^{-} \rightarrow e^{-} + e^{+} + e^{-}$, $\tau^- \rightarrow e^- +
\gamma$, $\tau^- \rightarrow \mu^- + \gamma$, $\mu^{-} + (A,Z)
\rightarrow e^{-} + (A,Z)$, $\etc$.  are allowed and should take place.
However, if the neutrino mixing in the weak charged lepton current,
eq.~(\ref{3numix}), is the only source of $L_e,~L_{\mu}$ and
$L_{\tau}$ non-conservation, as in the minimally extended Standard
Theory with massive neutrinos~\cite{SP76}, the rates and
cross sections of the indicated lepton flavour violating (LFV)
processes are suppressed by the factor~\cite{SP76} 
(see also~\cite{BPP77}) $(m_{j}/M_W)^4 < 6.7\times 10^{-43}$, $M_W$ being the
$W^{\pm}$ mass.  This renders them unobservable.  Indeed, the existing
upper limits on, $\eg$ $\mu \rightarrow e + \gamma$, $\mu \rightarrow
3e$, and $\tau \rightarrow \mu + \gamma$ decay branching ratios and on
the relative cross section of the $\mu - e$ conversion process
$\mu^{-} + \text{Ti} \rightarrow e^{-} + \text{Ti}$, are at least by 30
orders of magnitude bigger~\cite{mega,PDG04,BaBar05} (90\% C.L.):
\beq
\ba
\text{BR}(\mu \ra e+\gamma) < 1.2\times 10^{-11},~~~
\text{BR}(\mu \ra 3e) < 1.2\times 10^{-12}~,~\\[0.24cm]
\text{BR}(\tau \ra \mu +\gamma) < 6.8\times 10^{-8}~,~~~
\text{R}(\mu^{-} + \text{Ti} \rightarrow e^{-} + \text{Ti}) <
4.3\times 10^{-12}.  \ea \eeq
%
\noindent Future experiments with increased 
sensitivity can reduce the current bounds on $\text{BR}(\mu\ra
e+\gamma)$, $\text{BR}(\tau\ra \mu +\gamma)$ and on $\text{R}(\mu^{-}
+ (A,Z) \rightarrow e^{-} + (A,Z))$ by a few orders of magnitude (see,
$\eg$,~\cite{Kuno99}).  In the experiment MEG under preparation at 
PSI~\cite{psi} it is planned to reach a sensitivity to
\begin{equation}
 \text{BR}(\mu \ra e+\gamma)\sim (10^{-13} - 10^{-14})\,. 
\end{equation}
%
\indent It has been noticed in 1986 that in SUSY (GUT) theories with
see-saw mechanism of neutrino mass generation, the rates and cross
sections of the LFV processes can be strongly
enhanced~\cite{BorzMas86} and could be close to existing upper limits.
If the SUSY breaking occurs via soft terms with universal boundary
conditions at a scale $M_X$ above the RH Majorana neutrino mass scale
~\footnote{The possibility of ``flavour-blind'' SUSY breaking of
  interest is realised, $\eg$ in gravity-mediated SUSY breaking
  scenarios (see, $\eg$,~\cite{GMFB}).}  $M_R$, $M_{X}>M_R$, the
renormalisation group effects transmit the LFV from the neutrino
mixing at $M_{X}$ to the effective mass terms of the scalar leptons at
$M_R$, generating new LFV corrections to the flavour-diagonal mass
terms. {}For scalar lepton masses of $\sim$ few
hundred GeV, the LFV mass corrections at $M_R$ are typically of the
order of a few GeV and thus are much larger than the light neutrino
masses $m_j$.  As a consequence (and in contrast to the
non-supersymmetric case), the LFV scalar lepton mixing at $M_R$
generates additional contributions to the amplitudes of the LFV decays
and reactions which are not suppressed by the small values of neutrino
masses $m_j$.  As a result, the LFV processes can proceed with rates
and cross sections which are within the sensitivity of presently
operating and future planned experiments~\cite{BorzMas86,Hisano96}
(see also, $\eg$,~\cite{Iba01,JohnE,Saclay0105,PPY03,PPR3,PPTY03,Eiichi05}).

One of the basic ingredients of the see-saw mechanism is the matrix of
neutrino Yukawa couplings, $\mathbf{Y_{\nu}}$.  Leptogenesis depends
on $\mathbf{Y_{\nu}}$ as well.  In the large class of SUSY models with
see-saw mechanism of neutrino mass generation and flavour-universal
soft SUSY breaking at a scale $M_{X}>M_R$, we will consider,
the probabilities of LFV processes also depend strongly on
$\mathbf{Y_{\nu}}$ (see, $\eg$,~\cite{Iba01,JohnE}).  Thus, the matrix
of neutrino Yukawa couplings $\mathbf{Y_{\nu}}$ connects in the
see-saw theories the light neutrino mass generation with leptogenesis;
in SUSY theories with soft flavour-universal SUSY breaking at
$M_X>M_R$, $\mathbf{Y_{\nu}}$ links the light neutrino mass generation
and leptogenesis with LFV processes (see, $\eg$,~\cite{PPY03,PPR3}).

The matrix $\mathbf{Y_{\nu}}$ can be expressed in terms of the light
neutrino and heavy RH neutrino masses, the neutrino mixing matrix
$\pmns$, and an orthogonal matrix $\mathbf{R}$~\cite{Iba01}.
Leptogenesis can take place only if $\mathbf{R}$ is complex.
Obviously, $\mathbf{Y_{\nu}}$ depends on the Majorana phases in the
PMNS matrix $\pmns$. It was shown in~\cite{PPY03} (see 
also~\cite{JohnE}) that in the case of $\mathbf{R} \neq \mathbf{1}$, the
Majorana CP-violation phases in $\pmns$~\cite{BHP80} can affect
significantly the predictions for the rates of LFV decays $\mu
\rightarrow e + \gamma$, $\tau \rightarrow \mu + \gamma$, $\etc$.  in the
class of supersymmetric theories of interest.

The matrix $\mathbf{Y_{\nu}}$ can be defined, strictly
speaking, only at scales not smaller than $M_R$. The probabilities of
LFV processes depend on $\mathbf{Y_{\nu}}$ at the scale $M_R$,
$\mathbf{Y_{\nu}} =\mathbf{Y_{\nu}}(M_R)$.  In order to evaluate
$\mathbf{Y_{\nu}}(M_R)$ one has to know, in particular, the light
neutrino masses $m_j$ and the mixing matrix $\pmns$ at $M_R$, $\ie$,
one has to take into account the renormalisation group (RG)
``running'' of $m_j$ and $\pmns$ from the scale $M_Z \sim 100$ GeV, at
which the neutrino mixing parameters are measured, to the scale
$M_R$~\cite{Babu93,LolaE99,Casas99Maj,RGrunUEii,RGrunU}.

Working in the framework of the class of SUSY theories with see-saw
mechanism and flavour-universal soft SUSY breaking at a scale
$M_X>M_R$, we investigate in the present article the effects of RG
running of neutrino mixing parameters and of the Majorana CP-violation
phases on the predictions for the rates charged lepton flavour
violating decays.  We consider neutrino mass spectrum with normal
hierarchy (NH), $m_1 < m_2 < m_3$, allowing the lightest neutrino mass
$m_1$ to vary in the range $0 \leq m_1 \leq 0.30$ eV.  The latter
covers all possible varieties of light neutrino mass spectrum of the
NH type: {\it normal hierarchical, quasi-degenerate} as well as
spectrum with {\it partial normal hierarchy} (see,
$\eg$,~\cite{STPNu04}).  In order to find the changes of the neutrino
masses $m_j$ and mixing angles in $\pmns$ due to the RG evolution from
the scale $M_Z\sim 100$ GeV to $M_R$, the corresponding 1-loop RG
equations for the see-saw Majorana mass matrix of the left-handed
neutrinos were solved numerically.  The best fit values of the known
solar and atmospheric neutrino mixing angles and neutrino mass squared
differences at $M_Z$ were used as ``initial conditions'' for the
solution of the RG equations.  The values of the neutrino mixing
parameters at the scale $M_R$ served as input in the calculation of
$\mathbf{Y_{\nu}}(M_R)$ and correspondingly of the rates of the LFV
decays $l_i\to l_j + \gamma$.  The latter were performed for
quasi-degenerate heavy RH Majorana neutrinos for a number of typical
sets of values of the soft SUSY breaking parameters.  We have obtained
results both in the cases of $\mathbf{R} = \mathbf{1}$ and of complex
matrix $\mathbf{R}$.

Detailed predictions for the rates of the LFV processes in the class
of SUSY models with see-saw mechanism considered were
obtained, $\eg$ in 
refs.~\cite{Iba01,JohnE,Saclay0105,PPY03,PPR3,PPTY03,Eiichi05}.  However,
our study overlaps little with those performed in these articles.

%
\section{\large{Neutrino Mixing Parameters from Neutrino Oscillation Data}}
%
%
\hskip 1.0truecm
We will use the standard parametrisation of the
PMNS matrix $\pmns$ (see, $\eg$,~\cite{BPP1}): 
\bea 
\label{eq:Upara}
\pmns = \left( \bad 
 c_{12} c_{13} & s_{12} c_{13} & s_{13}e^{-i \delta}  \\[0.2cm] 
 -s_{12} c_{23} - c_{12} s_{23} s_{13} e^{i \delta} 
 & c_{12} c_{23} - s_{12} s_{23} s_{13} e^{i \delta} & s_{23} c_{13} 
\\[0.2cm] 
 s_{12} s_{23} - c_{12} c_{23} s_{13} e^{i \delta} & 
 - c_{12} s_{23} - s_{12} c_{23} s_{13} e^{i \delta} & c_{23} c_{13} 
\\ 
     \ea   \right) 
{\rm diag}(1, e^{i \frac{\alpha}{2}}, e^{i \frac{\beta_M}{2}}) \, ,
\eea
%
\noindent where 
$c_{ij} = \cos\theta_{ij}$, $s_{ij} = \sin\theta_{ij}$, the angles
$\theta_{ij} = [0,\pi/2]$, $\delta = [0,2\pi]$ is the Dirac
CP-violating phase and $\alpha$ and $\beta_M$ are two Majorana
CP-violation phases~\cite{BHP80,SchValle80D81}.  One can identify the
neutrino mass squared difference responsible for solar neutrino
oscillations, $\dmsol$, with $\Delta m^2_{21} \equiv m^2_2 - m^2_1$,
$\dmsol = \Delta m^2_{21} > 0$.  The neutrino mass squared difference
driving the dominant $\nu_{\mu} \rightarrow \nu_{\tau}$
($\bar{\nu}_{\mu} \rightarrow \bar{\nu}_{\tau}$) oscillations of
atmospheric $\nu_{\mu}$ ($\bar{\nu}_{\mu}$) is then given by
$|\dma|=|\Delta m^2_{31}|\cong |\Delta m^2_{32}| \gg \Delta m^2_{21}$.
The corresponding solar and atmospheric neutrino mixing angles,
$\theta_{\odot}$ and $\theta_{\rm A}$, coincide with $\theta_{12}$ and
$\theta_{23}$, respectively.  The angle $\theta_{13}$ is limited by
the data from the CHOOZ and Palo Verde experiments~\cite{CHOOZPV}.

The existing neutrino oscillation data allow us to determine $\Delta
m^2_{21}$, $|\Delta m^2_{31}|$, $\sin^2\theta_{12}$ and
$\sin^22\theta_{23}$ with a relatively good precision and to obtain
rather stringent limits on $\sin^2\theta_{13}$ (see,
$\eg$,~\cite{BCGPRKL2,3nuGlobal}).  The best fit values and the 95\%
C.L. allowed ranges of $\Delta m^2_{21}$, $\sin^2\theta_{12}$,
$|\Delta m^2_{31}|$ and $\sin^22\theta_{23}$ read
~\footnote{The data imply, in particular, that the high-LMA solution (see,
  $\eg$,~\cite{SNO3BCGPR}) is excluded at $\sim 3.3\sigma$.  Maximal
  solar neutrino mixing is ruled out at $\sim 6\sigma$; at 95\% C.L.\ 
  one finds $\cos 2\theta_\odot \geq 0.28$~\cite{BCGPRKL2}, which has
  important implications~\cite{PPSNO2bb}.}:
\beq
\label{bfvsol}
\ba
\deltasol = 8.0\times 10^{-5}~{\rm eV^2},~~
\sin^2\theta_{21} = 0.31~, \\[0.25cm]
\deltasol = (7.3 - 8.5) \times 10^{-5}~{\rm eV^2},~~
\sin^2 \theta_{12} = (0.26 - 0.36)~,
\ea
\eeq
\beq 
\label{eq:atmrange}
\ba
|\deltaatm| =2.2\times 10^{-3}~{\rm eV^2}~,~~\sin^22\theta_{23} = 1.0~, \\  [0.25cm]
|\deltaatm| = (1.7 - 2.9)\times 10^{-3}~{\rm eV^2}~,~~
\sin^22\theta_{23} \geq 0.90\;. 
\ea
\eeq
%
\noindent
A combined 3-$\nu$ oscillation
analysis of 
the solar neutrino, KL and CHOOZ data gives~\cite{BCGPRKL2,3nuGlobal}
\beq
\sin^2\theta_{13} < 0.027~(0.047),~~~~\mbox{at}~95\%~(99.73\%)~{\rm C.L.}
\label{th13}
\eeq
%
The neutrino oscillation parameters discussed above can (and very
likely will) be measured with much higher accuracy in the future (see,
$\eg$,~\cite{STPNu04}).

The sign of $\dma = \deltaatm $, as it is well known, cannot be
determined from the present (SK atmospheric neutrino and K2K) data.
The two possibilities, $\Delta m^2_{31(32)} > 0$ or $\Delta
m^2_{31(32)} < 0$ correspond to two different
types of $\nu$-mass spectrum:\\
-- {\it with normal hierarchy} 
$m_1 < m_2 < m_3$, $\dma=\Delta m^2_{31} >0$, and \\
-- {\it with inverted hierarchy} 
$m_3 < m_1 < m_2$, $\dma =\Delta m^2_{32}< 0$. \\
\noindent Depending on the sign of \dma, ${\rm sgn}(\dma)$, and 
the value of the lightest neutrino mass,
${\rm min}(m_j)$, the $\nu$-mass  spectrum can be\\
-- {\it Normal Hierarchical}: $m_1{\small \ll m_2 \ll }m_3$,
$m_2{\small \cong }(\dmsol)^ {1\over{2}}{\small \sim}$ 0.009 eV,
$m_3{\small \cong }|\dma|^{1\over{2}}{\small \sim}$ 0.045 eV;\\
-- {\it Inverted Hierarchical}: $m_3 \ll m_1 < m_2$,
with $m_{1,2} \cong |\dma|^{1\over{2}}\sim$ 0.045 eV; \\
-- {\it Quasi-Degenerate (QD)}: $m_1 \cong m_2 \cong m_3 \cong m_0$,
$m_j^2 \gg |\dma|$, $m_0 \gtap 0.10$~eV.

It is well-known, the theories employing the see-saw mechanism of
neutrino mass generation~\cite{seesaw} we are going to discuss
further, predict the massive neutrinos $\nu_j$ to be Majorana
particles.  Determining the nature of massive neutrinos is one of the
most formidable and pressing problems in today's neutrino physics.
Being of fundamental importance for understanding the origin of
$\nu$-masses and mixing, it is widely recognised as a highest priority
goal of the future studies of neutrino mixing 
(see, $\eg$,~\cite{STPNu04,APSbb0nu}).

If the massive neutrinos $\nu_j$ are Majorana fermions, getting
information about the Majorana CP-violation phases in $\pmns$ would be
highly problematic.  The oscillations of flavour neutrinos, $\nu_{l}
\rightarrow \nu_{l'}$ and $\bar{\nu}_{l} \rightarrow \bar{\nu}_{l'}$,
$l,l'=e,\mu,\tau$, are insensitive to the Majorana CP-violation phases
$\alpha$ and $\beta_M$~\cite{BHP80,Lang87}; they are insensitive to
the nature -- Dirac or Majorana, of massive neutrinos $\nu_j$.  The
only feasible experiments that at present have the potential of
establishing the Majorana nature of light neutrinos $\nu_j$ and of
providing information on the Majorana CP-violation phases in $\pmns$
are the experiments searching for the neutrinoless double beta
($\betabeta$-) decay, $(A,Z) \rightarrow (A,Z+2) + e^- + e^-$ (see,
$\eg$,~\cite{BiPet87,APSbb0nu,STPFocusNu04}).  
If the $\betabeta$-decay
is observed, the measurement of the $\betabeta$-decay half-life
combined with information on the absolute scale of neutrino masses (or
on ${\min}(m_j)$), might allow to significantly 
constrain, or possibly determine, the Majorana phase
$\alpha$ \cite{BPP1,BGKP96}, for instance. Establishing CP-violation
due to the Majorana phases in $\pmns$, however, was found to be
remarkably challenging~\cite{PPW}.

%
\section{\large{The See-Saw Mechanism and Neutrino Yukawa Couplings}}
%
%
\hskip 1.0truecm We consider the minimal supersymmetric standard model
with RH neutrinos (MSSMRN).  The superpotential for the lepton sector
is given by
\begin{align}
  \mathcal{W}_{l}=&\hat{E}^c_i(\mathbf{Y_{\rm E}})_{ij} \hat{L}_j\cdot
  \hat{H}_d + \hat{N}^c_i(\mathbf{Y_{\nu}})_{ij}\hat{L}_j\cdot
  \hat{H}_u - \frac{1}{2}(\mathbf{M_{\rm N}})_{ij}\hat{N}^c_i
  \hat{N}^c_j\;,
\label{eq_sptl}
\end{align}
%
where $\hat{L}_j$, $\hat{E}_{j }^{c}$ and $\hat{N}_{j}^c$,
$j=e,\mu,\tau \equiv 1,2,3$, are the chiral supermultiplets
respectively of the $SU(2)_L$ doublet lepton field $L_{jL}$ and of the
$SU(2)_L$ singlet charged lepton and neutrino fields $l_{j L}^{c}
\equiv C(\bar{l}_{j R})^{T}$ and $N_{jL}^c \equiv C\bar{N}_{j R}^{T}$,
where $l_{j R}$ and $N_{j R}$ are RH fields and $C$ is the charge
conjugation matrix, while $\hat{H}_u$ and $\hat{H}_d$ are the
supermultiplets of the two Higgs doublet fields $H_u$ and $H_d$
carrying weak hypercharges $-\frac 12$ and $\frac 12$.  In 
eq.~(\ref{eq_sptl}), $\mathbf{Y_{\rm E}}$ is the $3\times 3$ matrix of the
Yukawa couplings of the charged leptons, $\mathbf{Y_{\nu}}$ is the
$3\times 3$ matrix of neutrino Yukawa couplings, and $\mathbf{M_{\rm
    N}}$ is the Majorana mass matrix of the RH neutrinos $N_{j R}$.

In the framework of MSSMRN one can always choose a basis in which both
$\mathbf{Y_{\rm E}}$ and $\mathbf{M_{\rm N}}$ are diagonal.
Henceforth, we will work in that basis and will denote by 
$\mathbf{D_{\rm N}}$ the
corresponding diagonal RH neutrino mass matrix, $\mathbf{D_{\rm N}} =
{\rm diag}(M_1,M_2,M_3)$.

   Below the see-saw scale, $M_R = {\rm min}(M_j)$, the singlet
supermultiplets $\hat{N}^c_i$ containing the RH neutrino fields are
integrated out, and the superpotential, eq.~(\ref{eq_sptl}), becomes
\begin{align}
\mathcal{W}_{l}=&\hat{E}^c_i(\mathbf{Y_{\rm E}})_{ij}\hat{L}_j\cdot \hat{H}_d
-\frac{1}{2}(K_N)^{ij}(\hat{L}_i\cdot \hat{H}_u)(\hat{L}_j\cdot \hat{H}_u)\;,
\label{eq_sptl_bMR}
\end{align}
%
where
\begin{align}
(K_N)^{ij}=(Y_{\nu}^T)^{ik}(M_{\rm N}^{-1})^{kl}(Y_{\nu})^{lj}\;.
\label{eq_match_MR}
\end{align}
%
The last term yields Majorana mass term for the left-handed (LH)
flavour neutrinos after the electroweak symmetry breaking,
\begin{align}
\mathcal{L}_{m}^{\nu} = 
- \frac{1}{2}~\bar{\nu}^{C}_{Rj}~(m_{\nu})^{jk}~\nu_{Lk} + h.c.~,
\end{align}
%
where $\nu^{C}_{Rj} \equiv C (\bar{\nu}_{Lj})^{T}$ and
\begin{align}
(m_{\nu})^{ij}=(K_N)^{ij}\langle H_u\rangle^2\;.
\label{mnuKN}
\end{align}
%
Here $\langle H_u \rangle$ is the vacuum expectation value 
of $H_u$, $\langle H_u \rangle \equiv v_u = v \sin\beta$,
where $v = 174$ GeV.
The neutrino mass matrix is related to 
the light neutrino masses $m_j$ and the 
PMNS mixing matrix as
\begin{align}
(m_{\nu})^{ij}=(U^*)^{ik}m_k(U^{\dagger})^{kj}\;.
\label{mnuU}
\end{align}
%
 Using~(\ref{mnuKN}) and~(\ref{mnuU}), we can rewrite the matching condition,
eq.~(\ref{eq_match_MR}), at the energy scale $M_R$ in the form
\begin{align}
  \mathbf{U}^*(M_R)~\mathbf{D}_{\nu}(M_R)~ \mathbf{U}^{\dagger}(M_R) =
  - v_u^2~\mathbf{Y}_{\nu}^T(M_R)
  \mathbf{M}_{\rm N}^{-1}(M_R)\mathbf{Y}_{\nu}(M_R)\;.
\end{align}
%
where $\mathbf{D}_{\nu} = \mathrm{diag}(m_1,m_2,m_3)$.
Thus, in the basis in which the RH neutrino mass matrix
is diagonal $\mathbf{M}_{\rm N} = \mathbf{D}_{\rm N}$,
the neutrino Yukawa coupling at $M_R$ can be parametrised 
as~\cite{Iba01}
\begin{align}
  \mathbf{Y}_{\nu}(M_R) = \frac{1}{v_u}
  \sqrt{\mathbf{D}_N(M_R)}~\mathbf{R}~
  \sqrt{\mathbf{D}_{\nu}(M_R)}~\mathbf{U}^{\dagger}(M_R)\;.
\label{eq_para_yn}
\end{align}
%
Here $\mathbf{R}$ is a complex orthogonal matrix,
$\mathbf{R}^T\mathbf{R}= \mathbf{1}$.  The PMNS matrix $\mathbf{U}$
and the light neutrino masses in $\mathbf{D}_{\nu}$ in 
eq.~(\ref{eq_para_yn}) should be evaluated at $M_R$ with the help of the
renormalisation group equations which describe their evolution from
the the scale $M_Z$, where they are measured~\footnote{Strictly
  speaking, the neutrino mixing parameters are measured at scales
  $\sim 1$ GeV and $\sim 1$ MeV which are smaller than $M_Z$. However,
  these physical parameters do not get any significant corrections
  from the RG running below $M_Z$.}, to the scale $M_R$.

%
\section{\large{The Renormalisation Group Evolution Effects}}
\label{running}
%
%

\hskip 1.0truecm The RG running corrections to the neutrino mass
matrix in MSSM have been studied by many authors (see, 
$\eg$,~\cite{RGrunU,RGrunUEii} and the articles quoted therein).  In order to
find the RG evolution from $M_Z$ to $M_R$ of the neutrino mixing
angles $\theta_{12}$, $\theta_{23}$, $\theta_{13}$ and the masses
$m_j$, we have solved numerically the relevant 1-loop RG equation for
the neutrino mass matrix, eq.~(\ref{mnuKN}), in MSSMRN~\cite{Babu93}
(see also, $\eg$,~\cite{PPTY03,RGrunUEii,RGrunU}).  The case of neutrino
mass spectrum with normal hierarchy at $M_Z$, $m_1(M_Z) < m_2(M_Z) <
m_3(M_Z)$, was considered. The best fit values of the neutrino
oscillation parameters $\theta_{12}$, $\theta_{23}$, $\Delta m_{21}^2$
and $\Delta m_{31}^2$ at $M_Z$, given in eqs.~(\ref{bfvsol}) and
(\ref{eq:atmrange}), were used as input in the calculation of the RG
running effects.  The CHOOZ mixing angle at $M_Z$ was set to zero,
$\sin\theta_{13}(M_Z)=0$.

We will summarise below some of the results regarding the RG running
of the neutrino mixing parameters, which are relevant for our further
analysis.
\begin{itemize}
\item The corrections can become significant for relatively large
  $\tan\beta$.
\item The mixing angles are stable with respect to 
the RG running in the case of {\it normal hierarchical} 
neutrino mass spectrum, $m_1\ll m_2\ll m_3$.
\item For $m_1 \gtap 0.05$ eV, the RG running effects can change
  significantly the neutrino (lepton) mixing angles provided
  $\tan\beta$ is sufficiently large.
\item The solar neutrino mixing angle $\theta_{\odot} \equiv
  \theta_{12}$ at $M_R$ depends strongly on the Majorana phase
  $\alpha$, which plays very important role in the predictions
  of the effective Majorana mass in $\betabeta-$decay
  (see, e.g. \cite{STPFocusNu04}).  
  The effect of RG running for $\theta_{12}$ is largest 
  when $\alpha$ takes the CP-conserving value $\alpha = 0$;
  it is relatively small or negligible for the alternative 
  CP-conserving values of $\alpha$ \cite{LW81},
  $\alpha = \pm \pi$.  For $\alpha = 0$ and $\tan\beta \sim 50$, for
  instance, we have $\tan^2\theta_{12}(M_R) \ltap
  0.5\tan^2\theta_{12}(M_Z)$ for $m_1 \gtap 0.02$ eV.
\item The RG running effects depend weakly on the Majorana phase $\beta_M$.
\item The atmospheric neutrino mixing angle $\theta_{{\rm A}} \equiv
  \theta_{23}$ is essentially stable against the RG running
  corrections, except in the cases when $m_1$ and/or $\tan\beta$
  are/is relatively large, $\ie$,  (i) when $m_1 \gtap 0.2$ eV and
  $\alpha \ltap 3\pi/4$ if $\tan\beta \sim 10$, and (ii) for any $m_1$
  and $\alpha$ if $\tan\beta \gtap 40$.
\item The correction to $\sin\theta_{13}$ is relatively small even for
  large $\tan\beta$ if $m_1 \ltap 0.05$ eV,
  and for any $m_1 \gtap 0.30$ eV if $\theta_{13}(M_Z) \cong 0$ and
  $\alpha \cong 0$. For $\alpha = \pi$ and
  $\tan\beta \sim 50$ one can have $\sin\theta_{13}(M_R) \gtap 0.10$
  for $m_1 \gtap 0.08$ eV even if $\sin\theta_{13}(M_Z) = 0$.
  If, e.g. $\sin\theta_{13}(M_Z) = 0.10$ eV, 
  the RG effects lead for $\alpha = 0$ ($\pi/2 \ltap \alpha \ltap \pi$)
  to considerably smaller (larger) values of
  $\sin\theta_{13}(M_R)$ at $m_1 \gtap 0.05$ eV;
  for $\alpha \sim \pi/4$ one has
  $\sin\theta_{13}(M_Z) \cong \sin\theta_{13}(M_R)$. 
\item For sufficiently large $\tan\beta$, $\tan\beta \gtap 30$, the
  value of $\deltasol(M_R)$ depends strongly on $m_1$ in the interval
  $m_1 \gtap 0.05$ eV, and on $\alpha$ for $m_1 \gtap 0.10$ eV.  In
  the case of $\tan\beta \sim 50$, for instance, we have
  $\deltasol(M_R) \cong 3\deltasol(M_Z)$ ($\deltasol(M_R) \cong
  10\deltasol(M_Z)$) at $m_1 = 0.05~(0.10)$ eV.  The dependence of
  $\deltaatm(M_R)$ on $m_1$ and $\alpha$ is rather weak, unless
  $\tan\beta \gtap 40$ and $m_1 \gtap 0.10$ eV.
\item Some products of the neutrino mixing parameters, such as
  $s_{12}c_{12}c_{23}(m_1 - m_2e^{i \alpha})$ are practically stable
  with respect to RG running (for further discussion see
  Appendix~A and, $\eg$ refs.~\cite{PPTY03,RGrunUEii,RGrunU}).
\end{itemize}
%
%
Similar results hold also in the case of neutrino
mass spectrum with inverted hierarchy.

The RG equation for the neutrino mass matrix in MSSMRN and its
approximate solution as well as analytic expressions for some of the
neutrino mixing parameters at $M_R$ in the case of interest are given
in Appendix~A.  The specific features of the effects of RG running of
the neutrino mixing angles and mass (squared) differences discussed
above are illustrated in Figs.~\ref{m1-angle} and \ref{m1-mass},
where the values of $\tan^2\theta_{12}$, $\tan^2\theta_{23}$ and
$\sin\theta_{13}$, and of $(m_2 - m_1)$, $\deltasol$, $(m_3 - m_1)$
and $\deltaatm$, at $M_R = 2\times 10^{13}$ GeV are shown as functions
of ${\rm min}(m_j)=m_1$.  The neutrino mass spectrum is assumed to
be with normal hierarchy at $M_Z$.  The two figures correspond to the
following values of the neutrino oscillation parameters and $\beta_M$
at $M_Z$: $\tan^2\theta_{12} = 0.40$, $\tan^2\theta_{23} = 1$, $\Delta
m_{21}^2 = 8.0\times 10^{-5}~\text{eV}^2$, $\Delta m_{31}^2 =
2.2\times 10^{-3}~\text{eV}^2$, $\sin\theta_{13} = 0$ and $\beta_M=0$.

  The results obtained in the present Section are used in the
calculations of the $l_i \to l_j + \gamma$ decay 
rates to which we turn next.

%
\section{\large{Lepton Flavour Violation in MSSMRN and 
$l_i\rightarrow l_j + \gamma$ Decays}}
%
%
\hskip 1.0truecm
In the MSSMRN under discussion, the boundary conditions 
for the soft SUSY breaking are assumed to take place at some scale 
$M_X>M_R$, typically at $M_X\gtap M_{\rm GUT}$.
They are described in the general case by the Lagrangian  
\begin{align}
\mathcal{L}_{\text{soft}}=&-\frac{1}{2}M_{g3}
\bar{\tilde{G}}^{(a)c}\tilde{G}^{(a)}
-\frac{1}{2}M_{g2}\bar{\tilde{W}}^c_i\tilde{W}_i
-\frac{1}{2}M_{g1}\bar{\tilde{B}}^c\tilde{B}\nonumber\\
&
-(M_L^2)^{ij}\tilde{l}_{Li}^*\tilde{l}_{Lj}
-(M_E^2)^{ij}\tilde{e}_{Ri}^*\tilde{e}_{Rj}
-((A_E)^{ij}\tilde{e}_{Ri}^*\tilde{l}_{Lj}\cdot H_d+\text{h.c})\nonumber\\
&
-(m_{H1}^2|H_d|^2-m_{H2}^2|H_u|^2)
-(B\mu H_d\cdot H_u+\text{h.c.})
- \cdots
\label{soft1}
\end{align}
%
where $\tilde{G}^{(a)}$, $\tilde{W}_i$, and $\tilde{B}$ are the
gauginos associated with the $\text{SU}(3)_c$, $\text{SU}(2)_L$, and
$\text{U}(1)_Y$ gauge symmetry groups, and $H_u$, $H_d$,
$\tilde{l}_{Li}$, and $\tilde{e}_{Ri}^*$ are the scalar components of
$\hat{H}_u$, $\hat{H}_d$, $\hat{L}_i$, and $\hat{E}^c_i$ superfields,
respectively, and the dots mean additional soft RH sneutrino and
squark terms which do not play a role in our further discussion. We
will consider in what follows the class of MSSMRN in which the soft
SUSY breaking terms in eq. (\ref{soft1}) are flavour-universal 
at the scale $M_X$,
\begin{align}
  &M_{gi}=m_{1/2}\;,\nonumber\\
  &m_{H1}^2=m_{H2}^2=m_0^2\;, \nonumber\\
  &(M_L)^{ij}=(M_E)^{ij}=m_0^2~\delta^{ij}\;,\nonumber\\
  &(A_E)^{ij} = A_0(Y_E)^{ij}\;,~~A_0 = a_0 m_0\;, ~~(Y_E)^{ij}=Y_E^i~
  \delta^{ij}\;.
\end{align}
%
In this case the only lepton flavour mixing present at $M_X$ is that
due to the neutrino Yukawa couplings.  However, at the scale 
$M_R={\rm min}(M_j) = M_1$, where the RH neutrinos decouple, the RG
running effects induce flavour non-diagonal elements in the effective
soft SUSY breaking Lagrangian.  The off-diagonal elements of interest
for our further discussion are given in the leading-log approximation
by~\cite{Hisano96},
\begin{align}
\label{M2L}
(M_L^2)^{ij}\simeq& -\frac{1}{8\pi^2}m_0^2 
(3+|a_0|^2)
\sum_k (Y_{\nu}^{\dagger})^{ik}~\ln\frac{M_{X}}{M_k}~Y_{\nu}^{kj},~~i \neq j,\\
(M_E^2)^{ij} \simeq& ~0\;,~~i \neq j,\\
(A_E)^{ij}\simeq& -\frac{3}{8\pi^2}m_0 a_0 Y_E^i
\sum_k (Y_{\nu}^{\dagger})^{ik}~\ln\frac{M_{X}}{M_k}~Y_{\nu}^{kj}
\;,~~i \neq j.
\label{softlfv}
\end{align}
%
These terms generate new contributions in the amplitudes of the LFV
processes.  In the ``mass insertion'' 
(see, $\eg$,~\cite{Hisano96,JohnE,PPTY03})
and leading-log approximations,
the branching ratio of $l_i\to l_j + \gamma$ decay due to the new
contributions has the following form
\begin{align}
\text{BR}(l_i\to l_j\gamma)\simeq &
\frac{\Gamma(l_i\to e\nu\bar{\nu})}{\Gamma_{\text{total}}(l_i)}
\frac{\alpha_{\text{em}}^3}{G_F^2}
\frac{|(M_L^2)_{ij}|^2}{m_S^8}\tan^2\beta\nonumber\\
\simeq&
\frac{\Gamma(l_i\to e\nu\bar{\nu})}{\Gamma_{\text{total}}(l_i)}
\frac{\alpha_{\text{em}}^3}{G_F^2m_S^8}
\left|\frac{(3+a_0^2)m_0^2}{8\pi^2}\right|^2
\left|\sum_k (Y_{\nu}^{\dagger})^{ik}~\ln\frac{M_{X}}{M_k}~Y_{\nu}^{kj}
\right|^2\tan^2\beta\;,
\label{eq_ijg}
\end{align}
%
where $m_S$ represents SUSY particle mass.  It was 
shown in~\cite{PPTY03} that in most of the relevant soft 
SUSY breaking parameter space, the expression
\begin{align}
m_S^8\simeq 0.5~m_0^2~m_{1/2}^2~(m_0^2 + 0.6 ~m_{1/2}^2)^2\;,
\label{eq_ms}
\end{align}
%
gives an excellent approximation to the results obtained in a full
renormalisation group analysis, $\ie$, without using the leading-log
and the mass insertion approximations. Thus, in the case of soft SUSY
breaking mediation by flavour-universal terms at $M_X>M_R$, the
predicted rates of LFV processes such as $\mu\to e + \gamma$ decay are
very sensitive to the off-diagonal elements of
\begin{align}
\mathbf{Y}_{\nu}^{\dagger}(M_R)\mathbf{Y}_{\nu}(M_R)
= \frac{1}{v_u^2}~
\mathbf{U}(M_R)\sqrt{\mathbf{D}_{\nu}(M_R)}~\mathbf{R}^{\dagger}~
\mathbf{D}_N~\mathbf{R}~\sqrt{\mathbf{D}_{\nu}(M_R)}
\mathbf{U}^{\dagger}(M_R)\;.
\label{YnudYnu}
\end{align}
%
Obviously, $\mathbf{Y}_{\nu}^{\dagger}\mathbf{Y}_{\nu}$ is affected by
the RG running of $\mathbf{U}$ and $\mathbf{D}_{\nu}$ below the scale
$M_R$, which is strongly dependent on $\tan\beta$, the light neutrino
mass $m_1$ and the Majorana phase $\alpha$. 

\subsection{\large{Results for $\mathbf{R} = \mathbf{1}$}}

%

\hskip 1.0truecm In our further analysis we consider the case of
quasi-degenerate heavy RH neutrinos, $M_1=M_2=M_3 = M_R$ and we set
$M_X=M_{\rm GUT}=2\times 10^{16}$ GeV.  We also {\it first} fix
$\mathbf{R}$ in eq.~(\ref{eq_para_yn}) to be the unit matrix,
$\mathbf{R} = \mathbf{1}$. The case of $\mathbf{R} \neq \mathbf{1}$ is
discussed later.  Under the conditions specified above the branching
ratios $\text{BR}(l_i\to l_j + \gamma)$ depend on
$\mathbf{Y}_{\nu}^{\dagger}\mathbf{Y}_{\nu}$ which in turn takes the
form
\begin{align}
\mathbf{Y}_{\nu}^{\dagger}(M_R)\mathbf{Y}_{\nu}(M_R)=
\frac{M_R}{v_u^2} \mathbf{U}(M_R)
\mathrm{diag}(m_1(M_R), m_2(M_R), m_3(M_R))\mathbf{U}^{\dagger}(M_R)\;.
\label{ydagy}
\end{align}
%
{}For the off-diagonal elements of
$\mathbf{Y}_{\nu}^{\dagger}\mathbf{Y}_{\nu}$ of interest we get to
leading order in small parameters
\begin{subequations}
\begin{align}
(Y_{\nu}^{\dagger}Y_{\nu})_{12}\cong & ~\Delta_{21}c_{23}c_{12}s_{12}c_{13}
+\Delta_{31}s_{23}c_{13}s_{13}e^{-i\delta}+\mathcal{O}(\Delta_{21}s_{13})\; ,
\label{eq_mueg}
\\
(Y_{\nu}^{\dagger}Y_{\nu})_{13}\cong & ~
-\Delta_{21}s_{23}c_{12}s_{12}c_{13}
+\Delta_{31}c_{23}c_{13}s_{13}e^{-i\delta}+\mathcal{O}(\Delta_{21}s_{13})\;,
\label{eq_teg}
\\
(Y_{\nu}^{\dagger}Y_{\nu})_{23}\cong & ~
\Delta_{31}s_{23}c_{23}c_{13}^2+\mathcal{O}(\Delta_{21}, \Delta_{ji}s_{13},
s_{13}^2)\;,
\label{eq_tmug}
\end{align}
\end{subequations}
%
where all angles should
be evaluated at $M_R$ and 
\begin{align}
\Delta_{ij} \equiv
\frac{M_R}{v_u^2} \left(m_i(M_R)-m_j(M_R)\right) = 
\frac{M_R}{v_u^2}~\frac{\Delta m^2_{ij}(M_R)}{m_i(M_R) + m_j(M_R)}
\;.
\end{align}
%
\subsubsection{\large{The Case of Negligible RG Corrections}}
%
\hskip 1.0truecm The expressions for $(Y_{\nu}^{\dagger}Y_{\nu})_{kl}$, 
$k\neq l=1,2,3$, we have 
derived, eqs. (\ref{eq_mueg})-(\ref{eq_tmug}), 
are valid both for light neutrino mass spectrum 
with normal and inverted hierarchy, 
and, correspondingly, for NH, IH and QD spectrum.
If in the case of QD heavy RH neutrinos
the RG effects on the 
angles $\theta_{ij}$ and masses $m_j$ are
negligible or sufficiently small
($\eg$ if $\tan\beta \ltap 10$), 
in the large region 
of the soft SUSY breaking parameter space
where the approximate expression 
(\ref{eq_ijg}) with $m_S$ given by eq. (\ref{eq_ms})
reproduces with a relatively high precision 
the values of $\text{BR}(l_i\to l_j + \gamma)$
calculated exactly,
the ratios 
\begin{align}
\text{R}(ij/km) \equiv 
\frac{\text{BR}(l_i\to l_j + \gamma)}{\text{BR}(l_k\to l_m + \gamma)}\;,~~
i\neq j,k\neq m \;,
\end{align}
%
{\it depend only on the neutrino mixing parameters
$\theta_{12}$, $\theta_{23}$, $\theta_{13}$,
$\delta$ and $m_j$ at $M_Z$}. 
We have, for instance,
in the case of NH neutrino mass spectrum:
\begin{align}
\text{R}(21/31) \equiv \frac{\text{BR}(\mu \to e + \gamma)}
{\text{BR}(\tau \to e + \gamma)}
\cong \frac
{\left |\sqrt{\deltasol}c_{23} c_{12}s_{12}
+\sqrt{\deltaatm}s_{23}s_{13}e^{-i\delta} \right |^2}
{\left | 
\sqrt{\deltasol}s_{23} c_{12}s_{12}
-\sqrt{\deltaatm}c_{23}s_{13}e^{-i\delta} \right |^2}\;,~{\rm NH}\;.
\label{eq_2131NH}  
\end{align}
%
The expressions for IH (QD) spectrum can obtained
by replacing  $\sqrt{\deltaatm}$ with $(-2\sqrt{|\deltaatm|})$
($\sqrt{\deltasol}$ and $\sqrt{\deltaatm}$ with
$\deltasol$ and $\deltaatm$)
in eq. (\ref{eq_2131NH}). Depending on whether the terms 
$\propto \sqrt{\deltasol}$,  or the terms $\propto \sqrt{|\deltaatm|}$,
dominate, we get 
independently of the type of neutrino mass spectrum
\begin{align}
\text{R}(21/31)\cong \tan^{-2}\theta_{23}~{\rm or}~\tan^2\theta_{23}
\;.
\end{align}
%
In both cases $\text{BR}(\mu \to e + \gamma)$ 
and $\text{BR}(\tau \to e + \gamma)$ 
are predicted to differ at most by a factor $\sim 5$, which
reflects the current $3\sigma$ uncertainty in the value
of $\tan^2\theta_{23}$.

  Similarly, we have for the ratio $\text{R}(21/32)$:
\begin{align}
\text{R}(21/32)\equiv\frac{\text{BR}(\mu \to e + \gamma)}
{\text{BR}(\tau \to \mu + \gamma)}\cong 
(\deltaatm)^{-1}\left |\sqrt{\deltasol}c_{12}s_{12}s^{-1}_{23}
+\sqrt{\deltaatm}s_{13}c^{-1}_{23}e^{-i\delta} \right |^2\;,~{\rm NH}\;.
\label{eq_2132NH}
\end{align}
%
The expression for IH (QD) spectrum can be obtained
again by replacing  $\sqrt{\deltaatm}$ with $(-2\sqrt{|\deltaatm|})$
($\sqrt{\deltasol}$ and $\sqrt{\deltaatm}$ with
$\deltasol$ and $\deltaatm$) in eq. (\ref{eq_2132NH}). Now the 
dominance of the terms $\propto \sqrt{\deltasol}$ 
implies, $\eg$ for IH spectrum
\begin{align}
\text{R}(21/32)&\cong 
\frac{\deltasol}{4|\deltaatm|} \frac{c^2_{12}s^2_{12}}{s^2_{23}}\;,
~~{\rm IH}\;. 
\end{align}
%
In the case of NH spectrum this ratio is 4 times bigger, while
for QD spectrum the factor $(\deltasol/(4|\deltaatm|))$ 
should be replaced by $(\deltasol/|\deltaatm|)^2$.
If the terms $\propto \sqrt{|\deltaatm|}$ 
dominate, we have independently of 
the type of the spectrum: 
\begin{align}
\text{R}(21/32)\cong s^2_{13}c^{-2}_{23}\;.
\end{align}
%
In any of these cases we get 
$\text{BR}(\mu \to e + \gamma) \ll \text{BR}(\tau \to \mu + \gamma)$.

  Let us add that for the current best fit 
values of the neutrino oscillation parameters and 
$\sin^2\theta_{13}$ close to its current upper limit
(\ref{th13}), the terms $\propto \sqrt{\deltasol}$
and $\propto \sqrt{|\deltaatm|}$
in $\text{R}(21/31)$ and $\text{R}(21/32)$
are of the same order of magnitude 
for NH and IH neutrino mass spectra, 
while the term $\propto |\deltaatm|$
is the dominant one for QD spectrum.
Thus, if $\sin^2\theta_{13}$ is 
sufficiently large and the neutrino 
mass spectrum is of the NH or IH type,
the ratios $\text{R}(21/31)$ 
and $\text{R}(21/32)$ will 
depend on the Dirac CP-violation 
phase as well.

\subsubsection{\large{RG Effects}}

%

\hskip 1.0truecm  We turn next to the discussion of the RG effects.
In order to find the changes due to the RG running from $M_Z$
to $M_R$ of the neutrino mass matrix and correspondingly of the
neutrino mixing angles $\theta_{12}$, $\theta_{23}$, $\theta_{13}$ and
masses $m_j$, the 1-loop RG equations in MSSMRN for the matrix $K_N$
in eq.(10), as we have already discussed, are solved numerically. As
in the previous Section, the neutrino mass spectrum at $M_Z$ is
assumed to be with normal hierarchy, $m_1(M_Z) < m_2(M_Z) < m_3(M_Z)$.
The heavy RH neutrino mass scale $M_R$ is set at $M_R=2\times 10^{13}$
GeV. The same best fit values of the neutrino mixing parameters at
$M_Z$ were used as input in the calculation of the RG running effects.
In particular, we have fixed $\sin\theta_{13} = 0$ at $M_Z$.  The
values of the neutrino mixing parameters at the scale $M_R$ were used
as input in the calculation of the rates of the LFV decays $l_i\to l_j
+ \gamma$, which were performed for values of $\tan\beta$ from the
interval $\sim (10 - 50)$ and a number of typical sets of values of
the soft SUSY breaking parameters $m_0$, $m_{1/2}$ and $A_0$ in the
few 100 GeV range.  {\it These calculations have been performed
without employing the reading-log and the mass insertion
approximations.}

The results we have obtained using the methods described above show
that if $\mathbf{R} = \mathbf{1}$, taking into account the effects of
the RG running of the neutrino mixing parameters in the case of
relatively large $\tan\beta$ and/or $m_1$, can change significantly
the predictions for the LFV $l_i\to l_j + \gamma$ decay rates and the
corresponding branching ratios $\text{BR}(l_i\to l_j + \gamma)$.  The
magnitude of this change depends strongly on the value of the
Majorana CP-violation phase $\alpha$.  These conclusions are
illustrated in Fig.~\ref{m1-lfv}, where the predicted $l_i\to l_j +
\gamma$ decay branching ratios in the case of $\mathbf{R} =
\mathbf{1}$ are shown as functions of the light neutrino mass $m_1$
for two typical sets of soft SUSY breaking parameters and three values
of the Majorana CP-violation phase $\alpha$, $\alpha = 0;~\pi/2;~\pi$.

\begin{figure}
\begin{tabular}{ccc}
\includegraphics[scale=1.07]{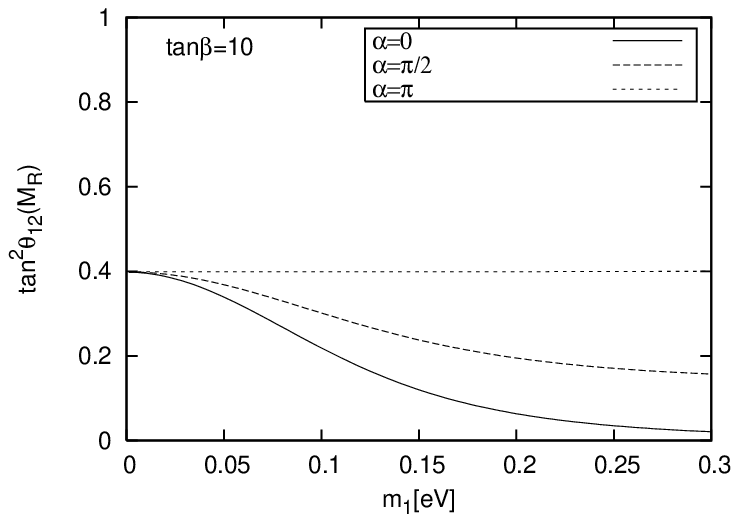}&
\includegraphics[scale=1.07]{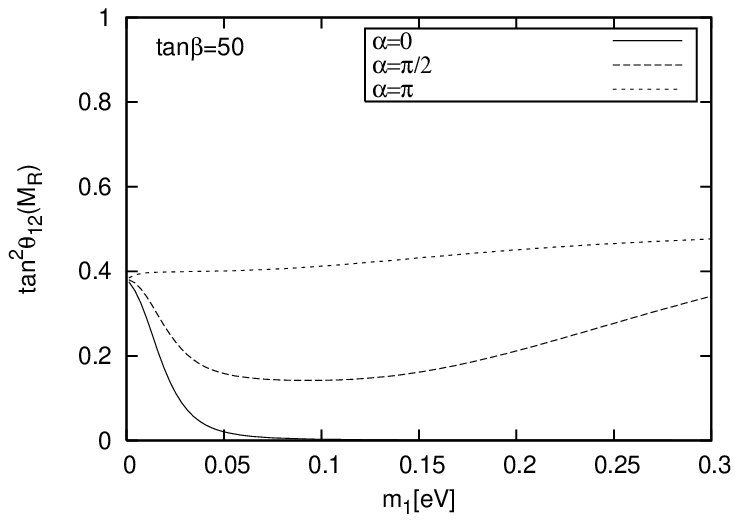}\\
\includegraphics[scale=1.07]{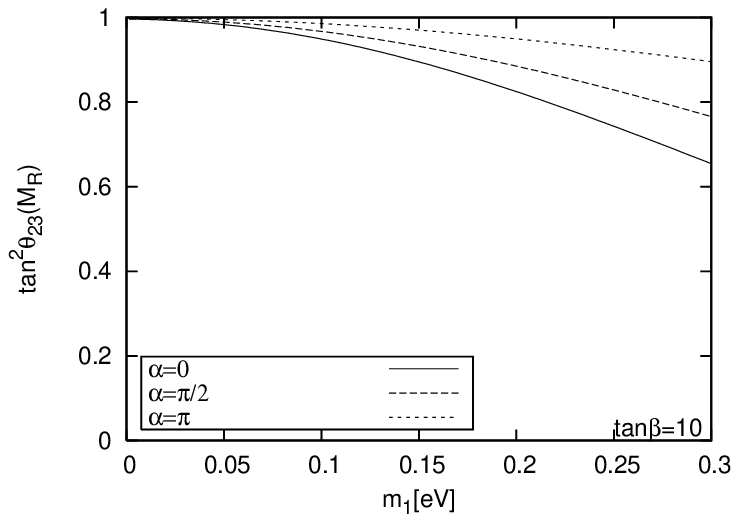}&
\includegraphics[scale=1.07]{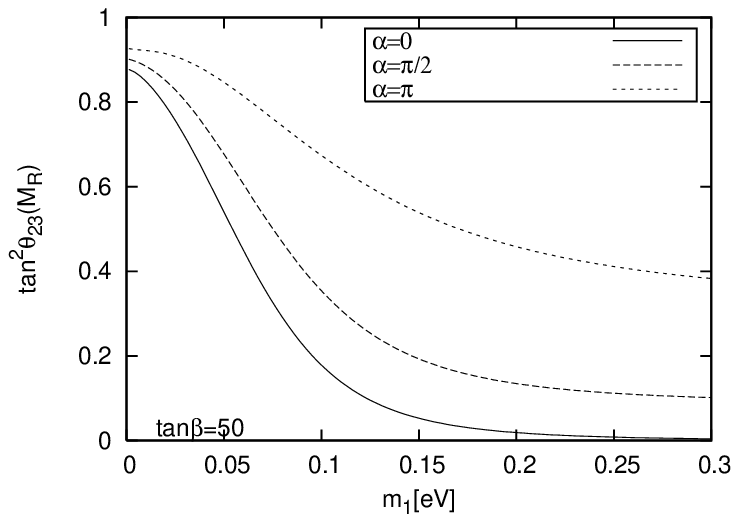}\\
\includegraphics[scale=1.07]{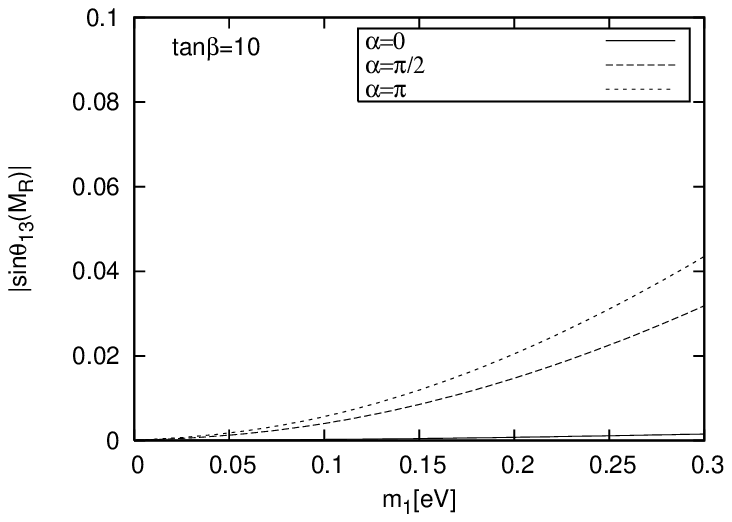}&
\includegraphics[scale=1.07]{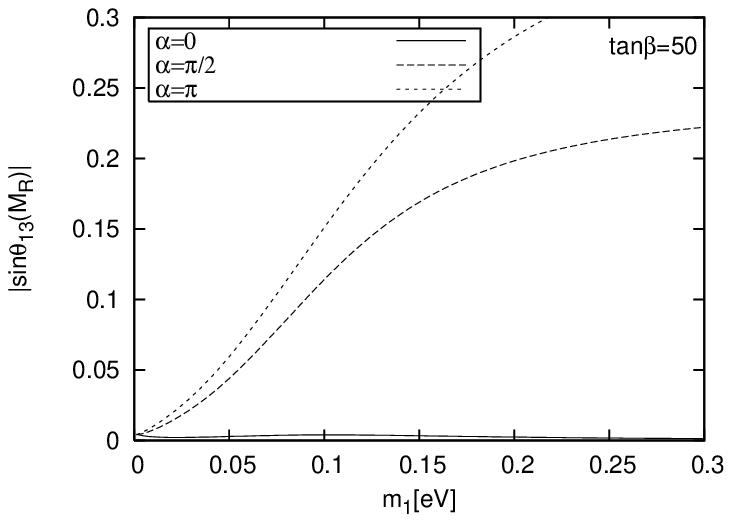}
\end{tabular}
\vspace{0.3cm}
\caption{The dependence of 
$\tan^2\theta_{\rm sol} \equiv \tan^2\theta_{12}$, 
$\tan^2\theta_{\rm atm} \equiv \tan^2\theta_{23}$ 
and of $|V_{13}| \equiv \sin\theta_{13}$, evaluated
at $M_R$, on the lightest neutrino 
mass $m_1$ for $\tan\beta = 10$ and $\tan\beta = 50$.
The following values neutrino mixing parameters at $M_Z$
were used as input in the calculation of the RG running effects:   
$\Delta m_{\odot}^2 = 8.0\times 10^{-5}~\text{eV}^2$, 
$\Delta m_A^2 = 2.2\times 10^{-3}~\text{eV}^2$, 
$\tan^2\theta_{\odot}=0.4$, $\tan^2\theta_A = 1$, and 
$\sin\theta_{13} = 0.0$. 
The neutrino mass spectrum at $M_Z$ is assumed
to be with normal hierarchy, 
$m_1(M_Z) < m_2(M_Z) < m_3(M_Z)$.
The RH neutrino mass scale 
$M_R$ is fixed as $M_R=2\times 10^{13}$~GeV.
}
\label{m1-angle}
\end{figure}

\begin{figure}
\begin{tabular}{cc}
\includegraphics[width=82mm,height=70mm]{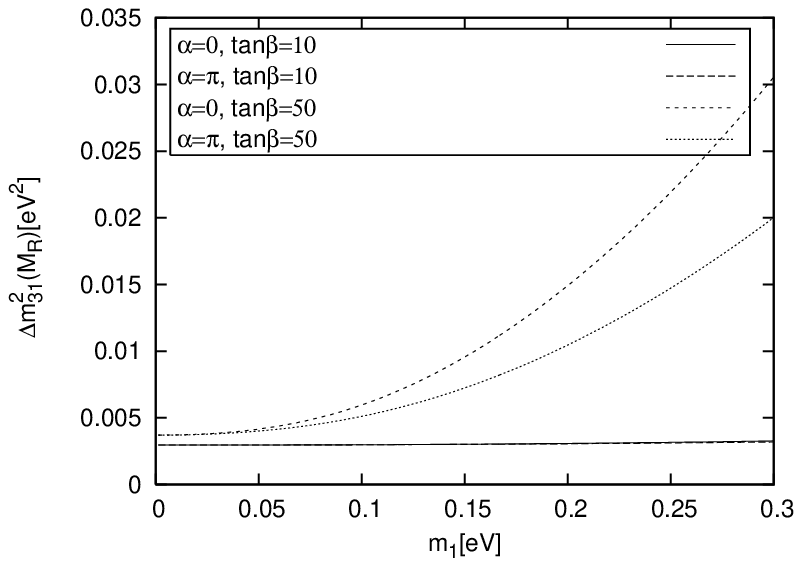}&
\includegraphics[width=82mm,height=70mm]{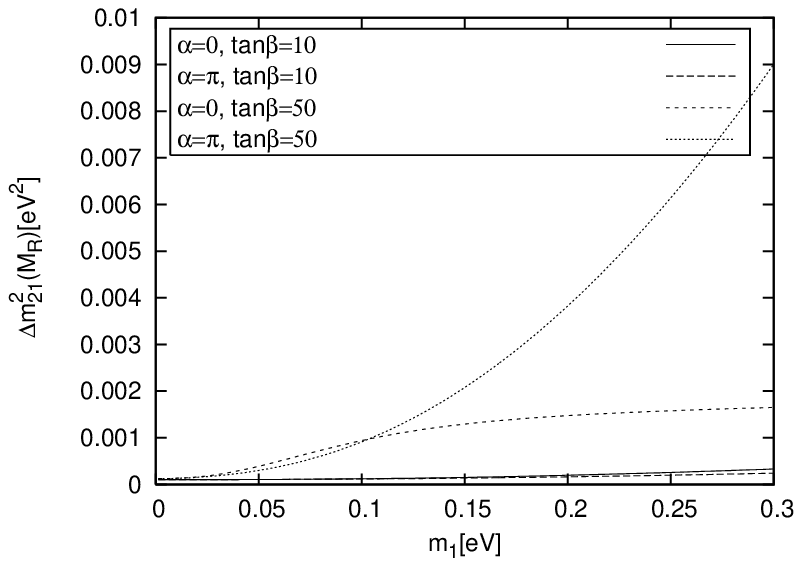}\\
(a)&(b)\\
\includegraphics[width=82mm,height=70mm]{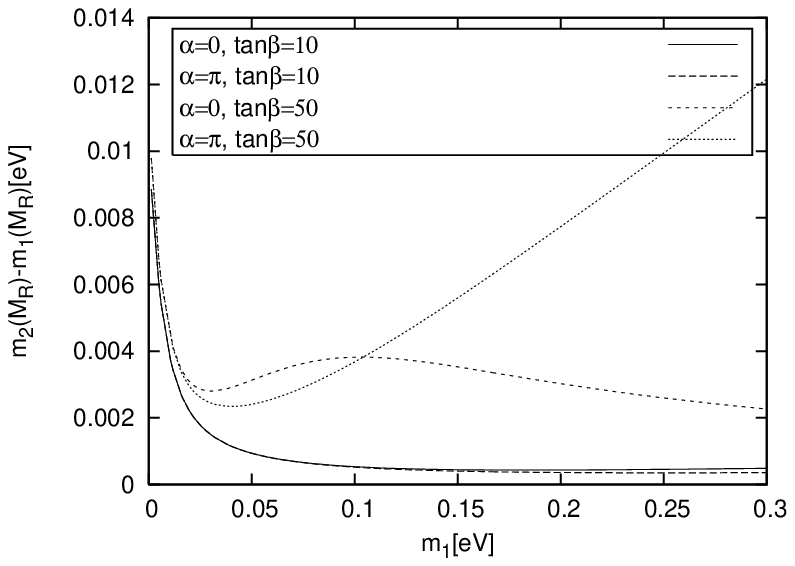}&
\includegraphics[width=82mm,height=70mm]{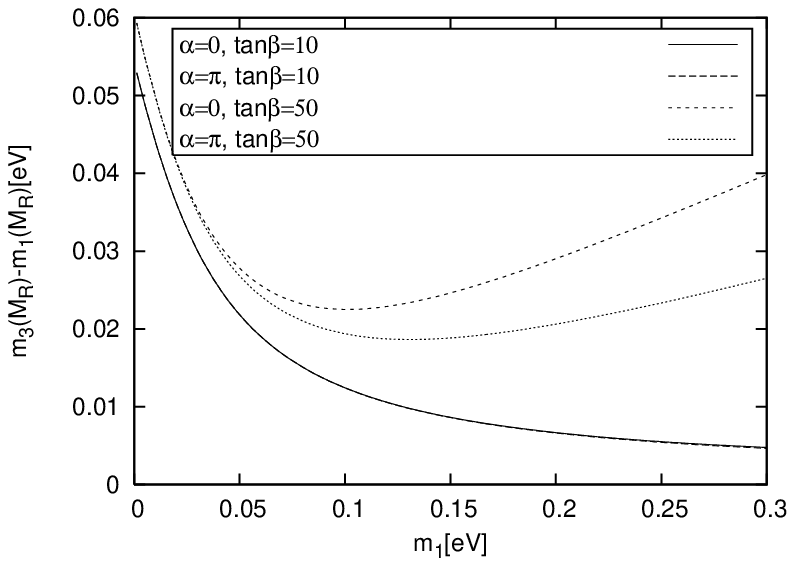}\\
(c)&(d)
\end{tabular}
\caption{The light neutrino mass differences 
at the scale $M_R$  as a functions of $m_1$
for $\alpha = 0; \pi$ and $\tan\beta=10;~50$:
(a) $\Delta m^2_A(M_R)\equiv m_3^2(M_R)-m_1^2(M_R)$, 
(b) $\Delta m^2_{\odot}(M_R)\equiv m_2^2(M_R)-m_1^2(M_R)$, 
(c) $m_2(M_R)-m_1(M_R)$, and (d) $m_3(M_R)-m_1(M_R)$. 
The neutrino mixing parameters at $M_Z$ and 
the mass $M_R$ used are the same as 
in Fig.~\ref{m1-angle}.
}
\label{m1-mass}
\end{figure}

We find, in particular, that for $ 0.05 \ltap m_1 \ltap 0.30$ eV and
$30 \ltap \tan\beta \ltap 50$, the predicted $ \mu \to e + \gamma$ and
and $\tau \to e + \gamma$ decay rates, or $\text{BR}(\mu \to e +
\gamma)$ and $\text{BR}(\tau \to e + \gamma)$, can be enhanced by the
effects of the RG running of $\theta_{ij}$ and $m_j$ by 1 to 3 orders
of magnitude if $\pi/4 \ltap \alpha \ltap \pi$, while $\text{BR}(\tau
\to \mu + \gamma)$ can be enhanced by up to a factor of 10.  The
enhancement of $\text{BR}(\mu \to e + \gamma)$ and $\text{BR}(\tau \to
e + \gamma)$ is due to the ``regeneration'' of the terms $\propto
\Delta_{31}s_{13}$ in $(Y_{\nu}^{\dagger}Y_{\nu})_{12}$ and
$(Y_{\nu}^{\dagger}Y_{\nu})_{13}$ at $M_R$: these terms are zero at
$M_Z$ since we have set $s_{13}(M_Z) = 0$.  The corresponding
enhancement factor in the case of $\mu \to e + \gamma$ and $\tau \to e
+ \gamma$ decays is given approximately by the ratio
\begin{align}
\label{EF}
F \cong \left (
\frac{\Delta m^2_{31}}{\Delta m^2_{21}}~
\frac{m_1 + m_2}{m_1 + m_3}~
\frac{s_{13}(M_R)}{c_{12}s_{12}}
\right )^2
\;,
\end{align}
where all quantities, and not only $s_{13}$, should be evaluated 
at $M_R$. For $\alpha = \pi$, however, $\theta_{12}(M_R) \cong
\theta_{12}(M_Z)$ (see Fig.~\ref{m1-angle}) and $F$ is reproduced with
a rather good approximation if one uses the values of $\Delta
m^2_{31}$, $\Delta m^2_{21}$ and of $m_j$ at $M_Z$. For $m_1 =
0.05~(0.10)$ eV, for instance, we have for $\tan\beta = 50$,
$s_{13}(M_R) \cong 0.05~(0.12)$ (see Fig.~\ref{m1-angle}) and for the
enhancement factor we find $F \cong 7~(60)$.  The case of $\tau \to e
+ \gamma$ decay is somewhat more complicated because of the
possibility of accidental mutual cancellation for $\alpha \cong \pi$
of the two terms in eq.~(\ref{eq_teg}) at $M_R$~\footnote{Note that
  since we set $s_{13} = 0$ at $M_Z$, such a cancellation can take
  place only due to the fact that the RG running can generate a
  non-zero and sufficiently large $s_{13}$ at $M_R$. }. This can lead
to a rather strong suppression (by more than an order of magnitude) of
$\text{BR}(\tau \to e + \gamma)$ in rather narrow interval of values
of $m_1$. The ``accidental'' suppression under discussion
can take place only in a limited region
of the soft SUSY breaking parameter space.

{}For $\alpha \cong 0$, the combined effect of the RG running of
$\theta_{12}$, $\theta_{23}$ $\theta_{13}$ and $m_j$ on
$\text{BR}(l_i\to l_j + \gamma)$ is relatively small even for $ m_1
\sim (0.2 - 0.3)$ eV and $\tan\beta \sim 50$.  This is a consequence
of the fact that for $\alpha = 0$, $s_{13}$ essentially does not
``run'' and $s_{13}(M_R) \cong s_{13}(M_Z)$, the effect of RG running
of $c_{23}$ ($s_{23}$) is relatively small, while the effects of RG
running of $\Delta_{21}$ compensates the effect of running of
$c_{12}s_{12}$ (see also Figs.~\ref{m1-angle} and \ref{m1-mass}).  
In this case $\text{BR}(\mu \to e
+ \gamma)$ and $\text{BR}(\tau \to e + \gamma)$ ($\text{BR}(\tau \to
\mu + \gamma)$) scale as functions of $m_1$ approximately as $\sim
[\Delta m^2_{21(31)}/(v_u(m_1 + m_{2(3)})]^2$, where $m_{2(3)} =
(m_1^2 + \Delta m^2_{21(31)})^{\frac{1}{2}}$ and the masses $m_j$ and
$\Delta m^2_{21,31}$ can be taken at $M_Z$.

The effects of the RG running of the neutrino mixing parameters
$\theta_{12}$, $\theta_{23}$, $\theta_{13}$ and $m_j$ are, in general,
considerably smaller for $\tan\beta \sim 10$.  They are practically
negligible for $\text{BR}(\tau \to \mu + \gamma)$.  If $\pi/2 \ltap
\alpha \ltap \pi$, they can lead to an enhancement of $\text{BR}(\mu
\to e + \gamma)$ approximately by a factor of 2.5 (20) for $m_1 =
0.10~(0.30)$ eV and are essentially negligible if $m_1 < 0.06$ eV.  In
what regards the $\tau \to e + \gamma$ decay branching ratio,
$\text{BR}(\tau \to e + \gamma)$, also in this case it can be strongly
suppressed for $\alpha \cong \pi$ due to the accidental cancellation
discussed above.  Such a suppression can be realised for relatively
large $m_1 \gtap 0.15$ eV only.

\subsection{\large{The Case of $\mathbf{R} \neq \mathbf{1}$}}
%

\hskip 1.0truecm We shall analyse next briefly the case of nontrivial
matrix $\mathbf{R}$: $\mathbf{R} \neq \mathbf{1}$.  Successful
leptogenesis can take place only if $\mathbf{R}$ is complex, so we
will consider $(\mathbf{R})^{*}\neq \mathbf{R}$.  The matrix
$\mathbf{R}$ of interest, being a complex orthogonal matrix, can be
written in the form: $\mathbf{R} = \mathbf{O}~e^{i\mathbf{A}}$, where
$\mathbf{O}$ is a {\it real orthogonal} matrix and $\mathbf{A}$ is a
{\it real antisymmetric} matrix, $(\mathbf{A})^T = - \mathbf{A}$.  In
the case of QD heavy RH Majorana neutrinos we can set $\mathbf{O} =
\mathbf{1}$ without loss of generality.  Thus, in the subsequent
calculations we use $\mathbf{R} = e^{i\mathbf{A}}$, parametrising the
real antisymmetric matrix $\mathbf{A}$ as follows
\begin{align}
A=
\begin{pmatrix}
0&a&b\\
-a&0&c\\
-b&-c&0
\end{pmatrix}\;,
\end{align}
%
where $a,b,c$ are real.  The following representation of
$e^{i\mathbf{A}}$ proves convenient for our analysis~\cite{PPY03}:
\begin{align}
\mathbf{R} = e^{i\mathbf{A}} = 
\mathbf{1}-\frac{\cosh r-1}{r^2}\mathbf{A}^2+i\frac{\sinh r}{r}\mathbf{A}\,,
\label{exp}
\end{align}
%
with $r=\sqrt{a^2+b^2+c^2}$.  The requirement of successful
leptogenesis in the case of QD light and heavy RH neutrino mass
spectra implies~\cite{PPY03} that $abc \neq 0$ and that $10^{-5}\ltap
|abc| \ll 1$. In what follows we will consider
the range of values of the parameters $a,b,c$
determined by $|a|,|b|,|c| \ltap 0.10$.

The matrix of neutrino Yukawa couplings in the case under discussion
has the form $\mathbf{Y}_{\nu}=\frac{\sqrt{M_R}}{v_u}
e^{i\mathbf{A}}\sqrt{\mathbf{D}_{\nu}}\mathbf{U}^{\dagger}$.  The
off-diagonal elements of $\mathbf{Y}_{\nu}^{\dagger}\mathbf{Y}_{\nu}$
of interest are given to leading order in the small quantities by
\begin{align}
\label{12R}
(Y_{\nu}^{\dagger}Y_{\nu})_{12}&= \Delta_{21}c_{23}c_{12}s_{12}c_{13}
+\Delta_{31}s_{23}c_{13}s_{13}e^{-i\delta}~~~~~~~~~~~~~~~~~~~~~~~~~~~~~~~~~~~~~~~~~~~~~~~~~~~~~~~~~~~~~~
\nonumber\\
&+2\frac{M_R}{v_u^2} i\biggl[ a\sqrt{m_1m_2} \left (
  c_{23}(c_{12}^2e^{-i\frac{\alpha}{2}}+
  s_{12}^2e^{i\frac{\alpha}{2}})
+ s_{13}c_{12}s_{12}s_{23}(e^{i(\frac{\alpha}{2}-\delta)} - 
e^{-i(\frac{\alpha}{2} +\delta)}) \right )
\nonumber\\
&\phantom{Space}
+b\sqrt{m_1m_3}s_{23} \left ( c_{12}e^{-i\frac{\beta_M}{2}}
 -s_{13}s_{12}e^{i(\frac{\beta_M}{2}-\delta)} \right )
\nonumber\\
&\phantom{Space}
+ c\sqrt{m_2m_3}s_{23} \left( s_{12}e^{i\frac{\alpha-\beta_M}{2}}
+ s_{13}c_{12}e^{-i(\frac{\alpha-\beta_M}{2} + \delta)} \right )
+\mathcal{O}(s^2_{13})
\biggr]+\mathcal{O}(r^2)\;,
\end{align}
\begin{align}
\label{13R}
(Y_{\nu}^{\dagger}Y_{\nu})_{13}&= -\Delta_{21}s_{23}c_{12}s_{12}c_{13}
+\Delta_{31}c_{23}c_{13}s_{13}e^{-i\delta}~~~~~~~~~~~~~~~~~~~~~~~~~~~~~~~~~~~~~~~~~~~~~~~~~~~~~~~~~~~~~~
\nonumber\\
&+2
\frac{M_R}{v_u^2} i\biggl[
a\sqrt{m_1m_2} \left ( -s_{23}(c_{12}^2e^{-i\frac{\alpha}{2}}+
s_{12}^2e^{i\frac{\alpha}{2}}) 
+ s_{13}c_{12}s_{12}c_{23}(e^{i(\frac{\alpha}{2}-\delta)} - 
e^{-i(\frac{\alpha}{2} +\delta)}) 
\right )
\nonumber\\
&\phantom{Space}
+b\sqrt{m_1m_3} \left ( c_{12}c_{23}e^{-i\frac{\beta_M}{2}}
 -s_{13}s_{12}s_{23}e^{i(\frac{\beta_M}{2}-\delta)} \right )
\nonumber\\
&\phantom{Space}
+ c\sqrt{m_2m_3}\left(s_{12}c_{23}e^{i\frac{\alpha-\beta_M}{2}}
+ s_{13}c_{12}s_{23} e^{-i(\frac{\alpha-\beta_M}{2} + \delta)} \right )
+\mathcal{O}(s^2_{13})
\biggr]+\mathcal{O}(r^2)\;,
\end{align}
\begin{align}
\label{23R}
(Y_{\nu}^{\dagger}Y_{\nu})_{23}&= \Delta_{31}s_{23}c_{23}c_{13}^2~~~~~~~~~~~~~~~~~~~~~~~~~~~~~~~~~~~~~~~~~~~~~~~~~~~~~~~~~~~~~~~~~~~~~~~~~
\nonumber\\
&+2\frac{M_R}{v_u^2}
i\biggl[
a\sqrt{m_1m_2} \left (-2i c_{12}s_{12}c_{23}s_{23}\sin\frac{\alpha}{2}
+ s_{13} \left [ c_{12}^2(s_{23}^2e^{-i(\frac{\alpha}{2}-\delta)} 
+ c_{23}^2e^{i(\frac{\alpha}{2} -\delta)}) \right. \right.
\nonumber\\
&\phantom{Space}
\left. \left.
+ s_{12}^2(s_{23}^2e^{i(\frac{\alpha}{2} +\delta)} 
+ c_{23}^2e^{-i(\frac{\alpha}{2} +\delta)})\right ]
\right )
\nonumber\\
&\phantom{Space}
+b\sqrt{m_1m_3} \left (-s_{12}(s_{23}^2e^{i\frac{\beta_M}{2}}
+c_{23}^2e^{-i\frac{\beta_M}{2}})
+s_{13}c_{12}c_{23}s_{23}~2i\sin(\frac{\beta_M}{2}-\delta)
\right )
\nonumber\\
&\phantom{Space}
+ c\sqrt{m_2m_3}\left ( c_{12}
(c_{23}^2e^{i\frac{\alpha-\beta_M}{2}}+
s_{23}^2e^{-i\frac{\alpha-\beta_M}{2}})
-s_{13}s_{12}c_{23}s_{23}~2i\sin(\frac{\alpha-\beta_M}{2} + \delta) 
\right )
\nonumber\\
&\phantom{Space}
+\mathcal{O}(s^2_{13}) \biggr]+\mathcal{O}(r^2)\;,
\end{align}
%
\hskip 1.0truecm  Equations (\ref{12R})-(\ref{23R})
are valid for any of the possible 
types of light neutrino mass spectrum.
The above results imply that
in the absence of significant 
RG effects, the ``double'' ratios
$\text{R}(21/31)=\text{BR}(\mu \to e + \gamma)
/\text{BR}(\tau \to e + \gamma)$ and
$\text{R}(21/32)=\text{BR}(\mu \to e + \gamma)/
\text{BR}(\tau \to \mu + \gamma)$, depend in the 
region of validity of eqs. (\ref{eq_ijg}) 
and (\ref{eq_ms}) in the relevant SUSY 
parameter space, {\it not only on 
$\theta_{ij}$, 
$\delta$ and $m_j$ at $M_Z$, 
but also on the Majorana CP-violation phases
$\alpha$ and $\beta_M$ and on
the ``leptogenesis parameters'' $a$, $b$ and $c$}.

\begin{figure}[t]
\begin{tabular}{cc}
\includegraphics[width=80mm, height=62mm]{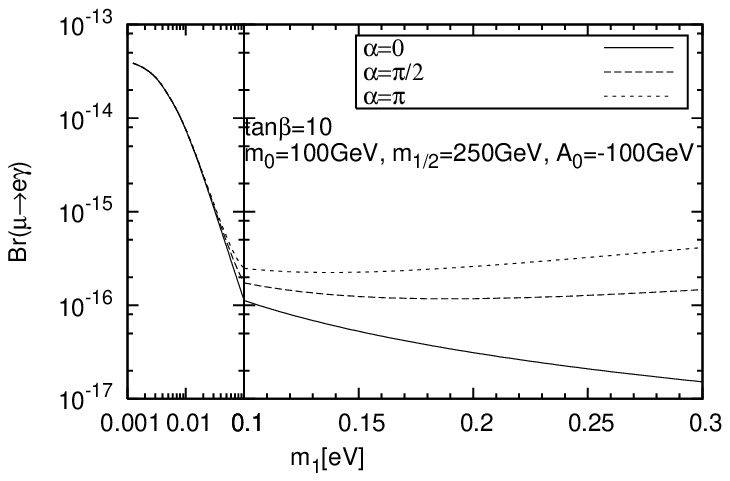}&
\includegraphics[width=80mm, height=62mm]{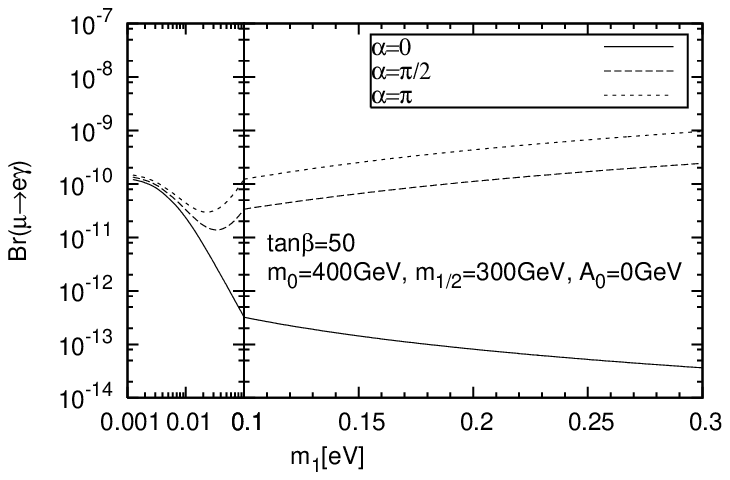}\\
\includegraphics[width=80mm, height=62mm]{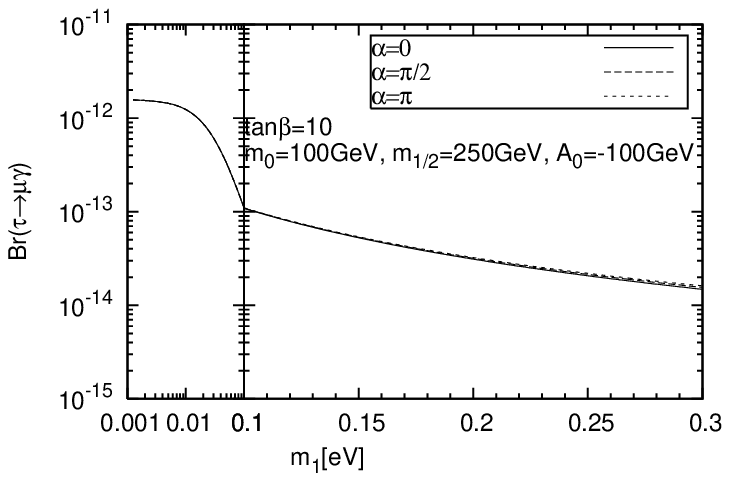}&
\includegraphics[width=80mm, height=62mm]{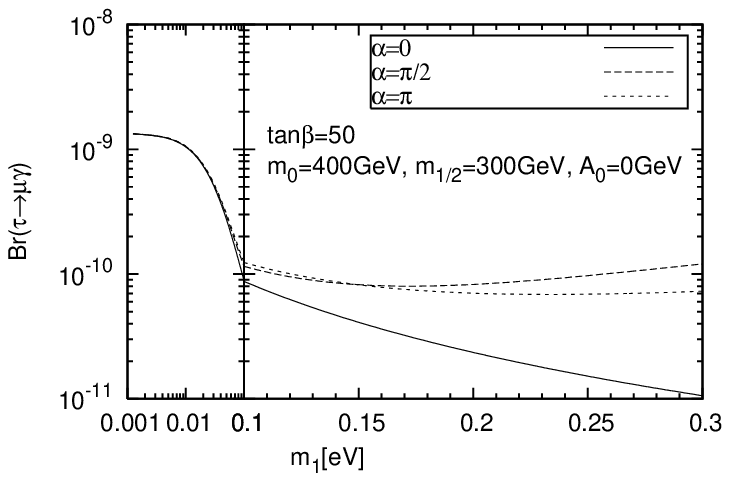}\\
\includegraphics[width=80mm, height=62mm]{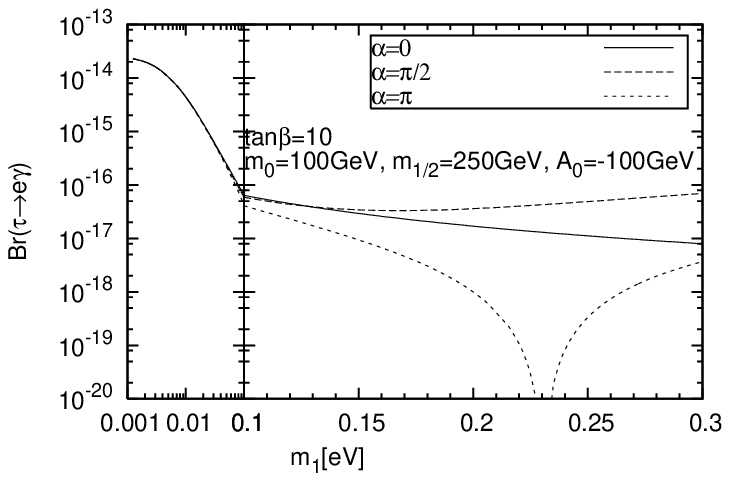}&
\includegraphics[width=80mm, height=62mm]{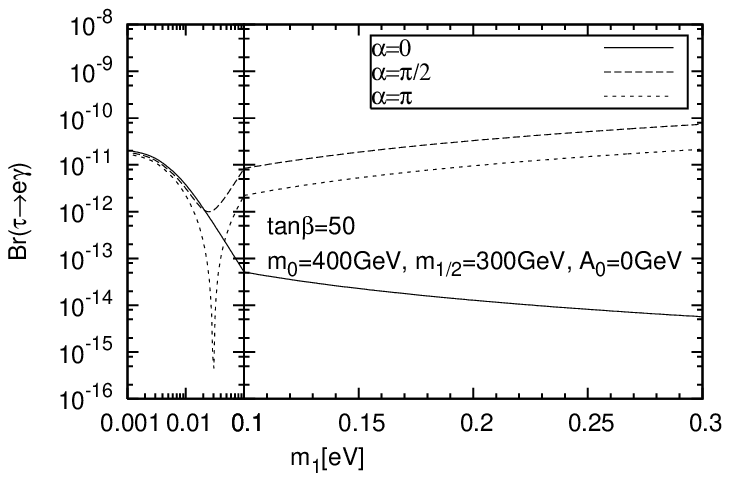}\\
(A)&(B)
\end{tabular}
\caption{The branching ratios of the LFV decays 
$\mu\to e + \gamma$, $ \tau \to \mu + \gamma$ and 
$\tau \to e + \gamma$ versus $m_1$ 
in the case of $\mathbf{R} = \mathbf{1}$ and for 
$\alpha = 0;~\pi/2;~ \pi$.     
The results shown correspond to the following two sets of SUSY parameters: 
(A) $\tan\beta=10$, $m_0=100$ GeV, $m_{1/2}=250$ GeV, $A_0=-100$ GeV, and
(B) $\tan\beta=50$, $m_0=400$ GeV, $m_{1/2}=300$ GeV, $A_0=0$.
The neutrino mixing parameters are the 
same as in Fig.~\ref{m1-angle}.
}
\label{m1-lfv}
\end{figure}

\begin{figure}
\begin{center}
\begin{tabular}{c}
\includegraphics[width=130mm,height=68mm]{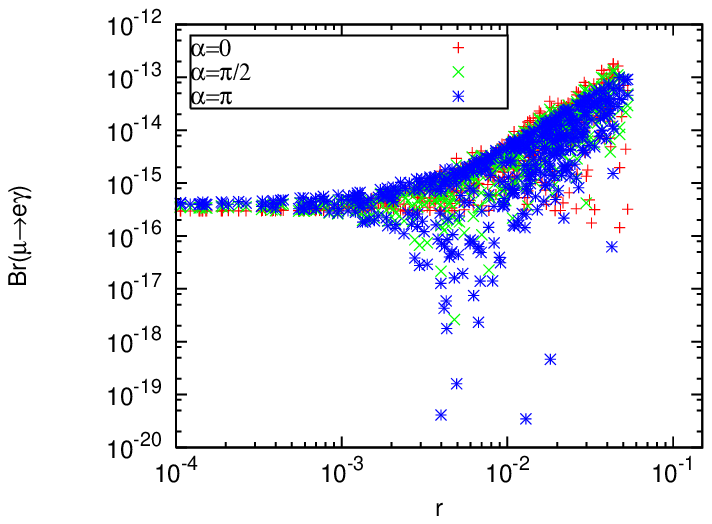}\\
\includegraphics[width=130mm,height=68mm]{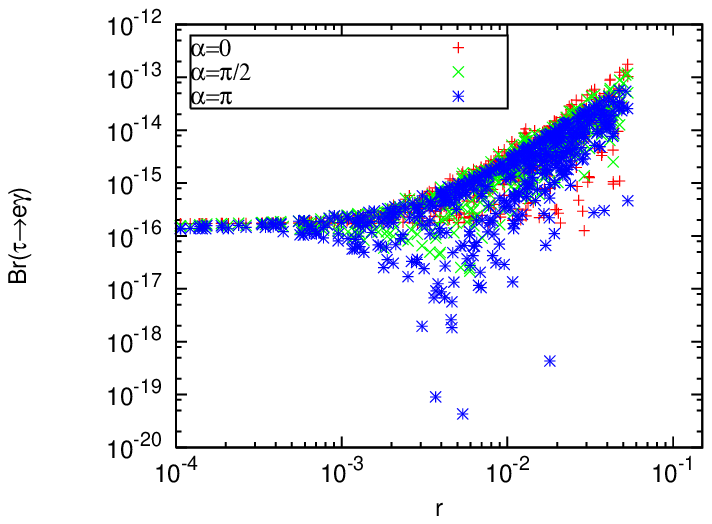}\\
\includegraphics[width=130mm,height=68mm]{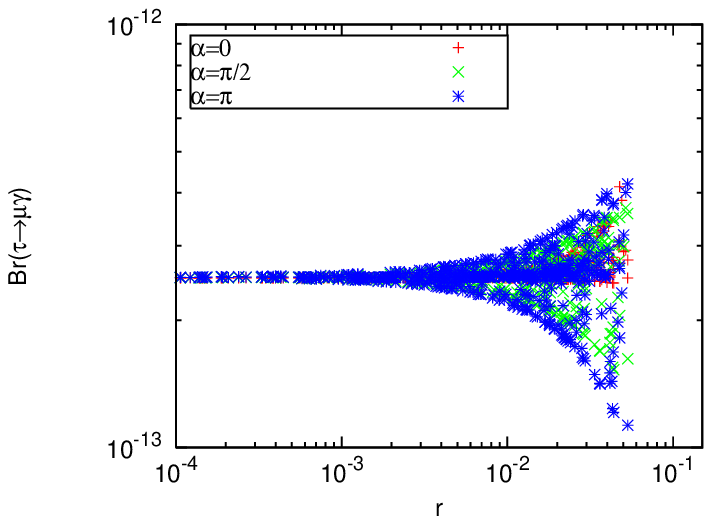}\\
\end{tabular}
\end{center}
\caption{The branching ratios
$\text{BR}(\mu\to e+\gamma)$, $\text{BR}(\tau\to e+\gamma)$,
and $\text{BR}(\tau\to \mu+\gamma)$ as functions
of $r\equiv \sqrt{a^2+b^2+c^2}$ for fixed  $m_1 = 0.06$ eV,
$\alpha = 0;~\pi/2;~\pi$ and $\beta_M = 0$.
The values of the other neutrino mixing parameters at
$M_Z$ are the same as in Fig.~\ref{m1-angle}.
The  SUSY parameters are taken as
$\tan\beta =1 0$, $m_{1/2}=250$ GeV, $m_0=100$ GeV and
$A_0 = -100$ GeV.
}
\label{m1-lfv-3}
\end{figure}

\begin{figure}
\begin{tabular}{cc}
\includegraphics[width=80mm, height=62mm]{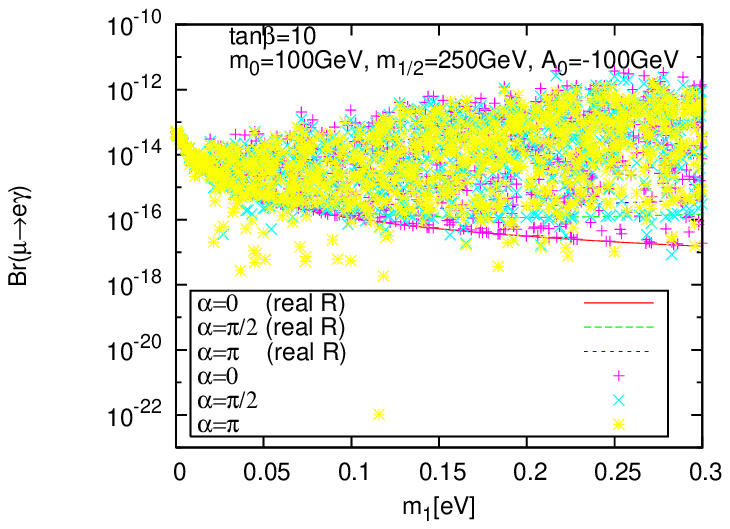}&
\includegraphics[width=80mm, height=62mm]{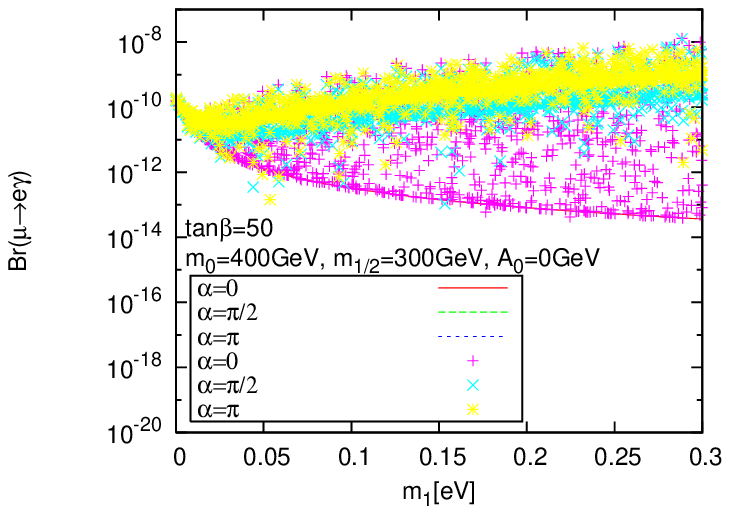}\\
\includegraphics[width=80mm, height=62mm]{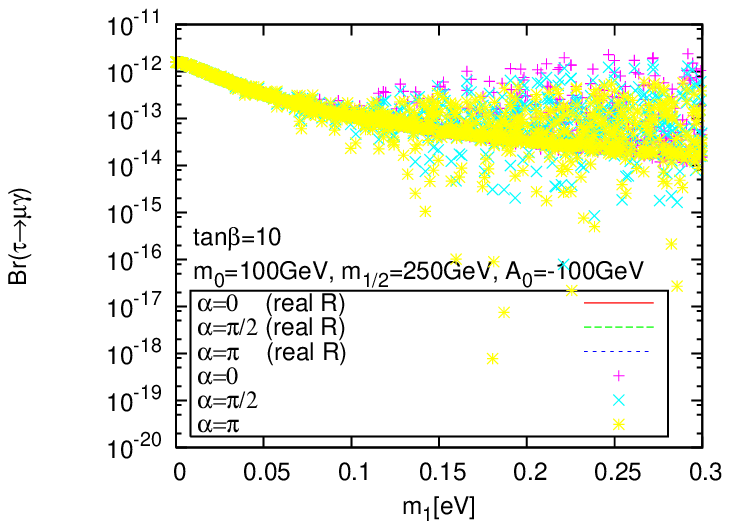}&
\includegraphics[width=80mm, height=62mm]{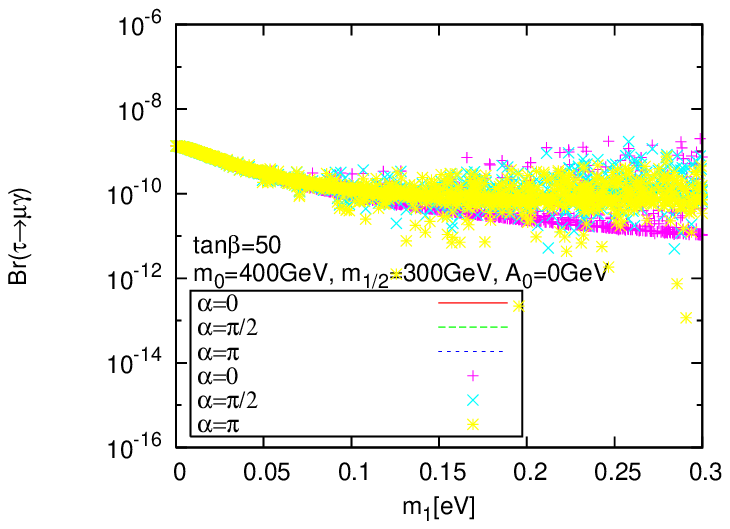}\\
\includegraphics[width=80mm, height=62mm]{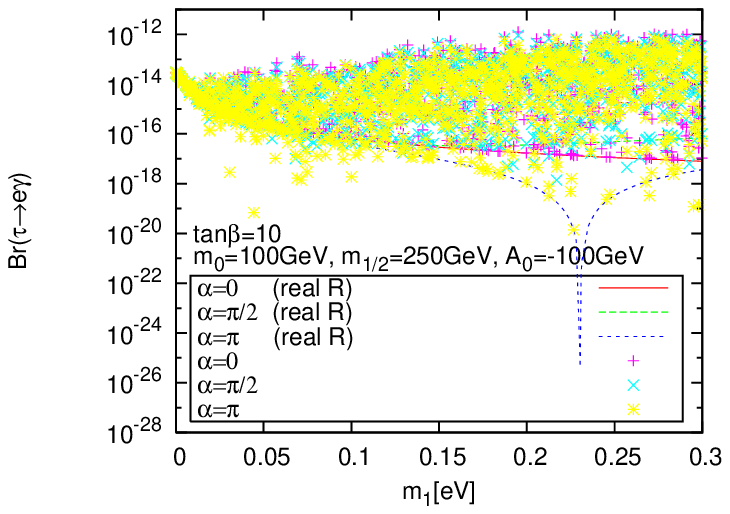}&
\includegraphics[width=80mm, height=62mm]{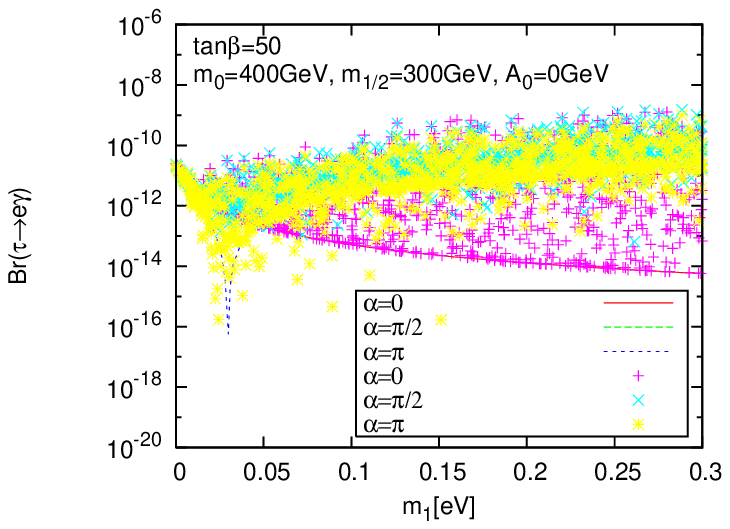}\\
(A)&(B)
\end{tabular}
\caption{The same as in Fig.~\ref{m1-lfv} but for 
  complex matrix $\mathbf{R} \neq \mathbf{1}$, eq.~(\ref{exp}).  The
  three constants $a,b,c$ parametrising the matrix $\mathbf{R}$ are
  assumed to lie in the interval $-0.1\leq a,b,c\leq 0.1$.  For
  comparison results for real $\mathbf{R} = \mathbf{1}$ are also shown
  (see text for further details).  }
\label{m1-lfv-2}
\end{figure}

  The results we find for $\text{BR}(l_i \to l_j + \gamma)$ 
taking into account the RG effects
in the case of complex $\mathbf{R}$ 
given by eq.~(\ref{exp}),  are illustrated in
{}Figs.~\ref{m1-lfv-3} and \ref{m1-lfv-2}. Both figures correspond to
$\beta_M=0$, $\alpha=0;~\pi/2;~\pi$, and the values of
$\theta_{12}$, $\theta_{23}$, $\theta_{13}$, $\Delta m^2_{21}$ and
$\Delta m^2_{31}$ at $M_Z$ used in Figs.~\ref{m1-angle}-\ref{m1-lfv}.  
In Fig.~\ref{m1-lfv-3}, the branching ratios
$\text{BR}(\mu\to e+\gamma)$, $\text{BR}(\tau\to e+\gamma)$, and
$\text{BR}(\tau\to \mu+\gamma)$ are shown as functions of $r\equiv
\sqrt{a^2+b^2+c^2}$ varying in the interval $10^{-4} - 5\times
10^{-2}$ for fixed $m_1 = 0.06$ eV and SUSY parameters $\tan\beta
=10$, $m_{1/2}=250$ GeV, $m_0=100$ GeV and $A_0 = -100$ GeV.
{}Figure~\ref{m1-lfv-2} is similar to Fig.~\ref{m1-lfv}, but shows
results for complex $\mathbf{R}$. It is obtained assuming that the
three constants $a,b,c$ parametrising the matrix $\mathbf{R}$ lie in
the interval $-0.1\leq a,b,c\leq 0.1$.  For comparison, in
Fig.~\ref{m1-lfv-2} results for $\mathbf{R} = \mathbf{1}$ are also
shown~\footnote{ We should note that some of the predicted values of
the branching ratio $\text{BR}(\mu\to e\gamma)$, presented in
Figs.~\ref{m1-lfv} and \ref{m1-lfv-2}, are already ruled out by the
existing experimental limits on $\text{BR}(\mu\to e\gamma)$. This
illustrates, in particular, the type of constraints one might obtain, 
$\eg$ on the Majorana phase $\alpha$ and/or $m_1$ for a given specific 
set of values of SUSY breaking parameters.}.

In the case of $\tan\beta \sim 10$ and $m_1 \sim 0.06$ eV 
(Fig.~\ref{m1-lfv-3}), 
the RG effects are negligible.  For $\alpha = 0$, the
branching ratios $\text{BR}(\mu\to e+\gamma)$ and $\text{BR}(\tau\to
e+\gamma)$ can be significantly enhanced by the effect of complex
$\mathbf{R}$~\cite{PPY03} for $r \gtap 5\times 10^{-3}$: the
enhancement factor can be as large as $\sim 100$ at $r \cong 5\times
10^{-2}$.  Similar enhancement can take place for values of $\alpha$
in the interval $[\pi/2,\pi]$. However, as Fig.~\ref{m1-lfv-3} shows,
{\it for these relatively large values of $\alpha$, $\text{BR}(\mu\to
  e+\gamma)$ and $\text{BR}(\tau\to e+\gamma)$ can also be suppressed
  as a consequence of a partial cancellation between the contributions
  due to the complex $\mathbf{R}$ ($r \neq 0$) and the contribution
  present for $\mathbf{R} = \mathbf{1}$} 
(see eqs.~(\ref{12R})-(\ref{23R})).  For the chosen values $m_1$ and
$\beta_M$, the effect of complex $\mathbf{R}$ on $\text{BR}(\tau\to
\mu+\gamma)$ is minor: the latter can change at most by a factor $\sim
2$ for $r \ltap 5\times 10^{-2}$.

In the case of $|a|,|b|,|c| \leq 0.10$ and $\tan\beta = 10;~50$,
illustrated in Fig.~\ref{m1-lfv-2}, $r$ can take somewhat larger
values.  For $\tan\beta = 10$, the contributions due to the matrix
$\mathbf{R}$ in $\text{BR}(\mu\to e+\gamma)$ and $\text{BR}(\tau\to
e+\gamma)$ become dominant at $m_1 \gtap 0.01$ eV; for
$\text{BR}(\tau\to \mu +\gamma)$ the same is valid at $m_1 \gtap 0.10$
eV.  Correspondingly, for $m_1 \gtap 0.01$ eV ($m_1 \gtap 0.10$ eV),
$\text{BR}(\mu\to e+\gamma)$ and $\text{BR}(\tau\to e+\gamma)$
($\text{BR}(\tau\to \mu +\gamma)$) can be enhanced significantly
independently of the value of the Majorana phase 
$\alpha$ (Fig.~\ref{m1-lfv-2}).  For relatively large values of $\alpha$, $\eg$
$\alpha \sim [\pi/2,\pi]$, there is again the possibility of a
reduction (or suppression) of the $l_i \to l_j + \gamma$ decay rates
of interest with respect to those predicted for $\mathbf{R} =
\mathbf{1}$: the presence of the phase $\alpha$ can lead to partial
cancellations between the three new terms proportional to
$\sqrt{m_1m_2}$, $\sqrt{m_1m_3}$ and $\sqrt{m_2m_3}$ in eqs.~(\ref{12R})-(\ref{23R}).  Similar 
conclusions are valid in the case
of $\tan\beta = 50$.  For $\pi/2 \ltap \alpha \leq \pi$ in this case,
the enhancement of $\text{BR}(\mu\to e+\gamma)$ and $\text{BR}(\tau\to
e+\gamma)$ at $m_1 \gtap 0.02$ eV due to the RG effects is so strong
that the contributions due to the complex $\mathbf{R}$ can lead to
further increase of $\text{BR}(\mu\to e+\gamma)$ and
$\text{BR}(\tau\to e+\gamma)$ by not more than approximately an order
of magnitude. For sufficiently large $\alpha$, $\eg$ $\alpha \sim
[\pi/2,\pi]$, these contributions can also compensate partially the RG
effect of enhancement leading to significantly smaller
$\text{BR}(\mu\to e+\gamma)$ and $\text{BR}(\tau\to e+\gamma)$ than in
the case of $\mathbf{R} = \mathbf{1}$ (Fig.~\ref{m1-lfv-2}).
Qualitatively similar results are valid also for $\text{BR}(\tau\to
\mu +\gamma)$.

  In the preceding discussion we have focused on the role of
the Majorana phase $\alpha$, ignoring the possible effects
of the second Majorana phase in the PMNS matrix $\beta_M$. 
Figures~\ref{m1-lfv-3} and \ref{m1-lfv-2}
correspond to $\beta_M = 0$. It follows from
eqs.~(\ref{12R})-(\ref{23R}) that if the phase
$\beta_M \neq 0$ is sufficiently large,
it can have an effect on the contributions
to $\text{BR}(l_i \to l_j + \gamma)$
due to the complex matrix $\mathbf{R}$, eq.~(\ref{exp}),
similar to that of the phase $\alpha$ discussed above.

%
\section{\large{Conclusions}}
%
%
\hskip 1.0truecm In the present article we have investigate the
effects of the Majorana CP-violation phases in the PMNS matrix and of
the RG evolution of the light neutrino mixing parameters -- masses and
PMNS mixing angles, on the predictions for the rates of lepton flavour
violating (LFV) decays, $\mu\to e+\gamma$, $\tau\to e + \gamma$ and
$\tau \to \mu + \gamma$, in a class of SUSY theories with see-saw
mechanism of neutrino mass generation.  We have considered minimal
supersymmetric extensions of the Standard Theory with heavy RH
Majorana neutrinos (MSSMRN), in which soft SUSY breaking by 
flavour-universal terms in the Lagrangian occurs at a scale $M_{\rm GUT}$ 
above the RH Majorana neutrino mass scale $M_R$. 
In this class of theories, the charged lepton 
radiative decays $\mu\to e+\gamma$, $\tau \to \mu +
\gamma$, $\etc$.  and other LFV processes are predicted to take place
with rates that typically are within the reach of present and planned
experiments.  The predictions for the $\mu\to e+\gamma$, $\tau \to \mu
+ \gamma$, $\etc$.  decay rates are known to depend critically on the
matrix of neutrino Yukawa couplings at the scale $M_R$,
$\mathbf{Y_{\nu}}(M_R)$.  The matrix $\mathbf{Y_{\nu}}(M_R)$ can be
expressed in terms of the light and heavy RH Majorana neutrino masses,
the PMNS mixing matrix $U$ -- all at the scale $M_R$, and an
orthogonal matrix $\mathbf{R}$.  The same matrix
$\mathbf{Y_{\nu}}(M_R)$ is one of the basic ingredients of the see-saw
mechanism and of the leptogenesis which can take place only if the
orthogonal matrix $\mathbf{R}$ is complex.

The neutrino mixing parameters entering into the expression for
$\mathbf{Y_{\nu}}(M_R)$ should be evaluated at the scale $M_R$.  These
include the solar and atmospheric neutrino mixing angles $\theta_{12}$
and $\theta_{23}$, the CHOOZ angle $\theta_{13}$, the two Majorana
CP-violating phases in the PMNS matrix $\alpha$ and $\beta_M$ 
(see eq.~(\ref{eq:Upara})), and the light neutrino masses $m_1$, $m_2 =
\sqrt{m_1^2 + \Delta m^2_{21}}$, and $m_3 = \sqrt{m_1^2 + \Delta
  m^2_{31}}$, where $\Delta m^2_{21}$ and $\Delta m^2_{31}$ are the
neutrino mass squared differences driving the solar and atmospheric
neutrino oscillations.  In order to find the changes of the mixing
angles $\theta_{ij}$ and masses $m_j$, due to the RG evolution from
the scale $M_Z\sim 100$ GeV, where they are measured, to $M_R$, we
have solved numerically the 1-loop RG equations in MSSMRN for the
see-saw Majorana mass matrix of the left-handed flavour neutrinos, eq.
(\ref{mnuKN}).  The light neutrino mass spectrum was assumed to be
with normal hierarchy, $m_1 < m_2 < m_3$, and the lightest neutrino
mass $m_1$ was allowed to take values from 0 to 0.30 eV.  The current
best fit values of the known neutrino mixing angles and $\Delta
m_{21(31)}^2$ at $M_Z$ (see eqs. (\ref{bfvsol}) and
(\ref{eq:atmrange})) were used as initial conditions 
for the solution of the RG equations. We have presented results 
in the case of the CHOOZ mixing angle $\theta_{13}$ and the
Majorana phase $\beta_M$  set to 0 at $M_Z$.  In these and the
subsequent calculations the scale $M_R$ was fixed at $M_R = 2\times
10^{13}$ GeV.  Our results concerning the RG evolution of the neutrino
mixing angles and neutrino mass differences $(m_{2(3)} - m_1)$ and
$\Delta m^2_{21(31)}$ are discussed in Section \ref{running} and are
illustrated in Figs. \ref{m1-angle} and \ref{m1-mass}, where the
indicated quantities are shown at the scale $M_R$ as functions of
$m_1$.

The values of the neutrino mixing parameters at the scale $M_R$ were
used as input in the calculation of $\mathbf{Y_{\nu}}(M_R)$ and
correspondingly of the rates of the LFV decays $l_i\to l_j + \gamma$.
The latter were performed for a number of typical sets of values of
the soft SUSY breaking parameters ($m_0$, $m_{1/2}$, $A_0$) in the few
100 GeV range and for values of $\tan\beta$ from the interval (10-50).
{}For simplicity, the analyses were performed for the case of
quasi-degenerate in mass heavy RH Majorana neutrinos,
$M_1=M_2=M_3=M_R$.  
The rates of the LFV
decays $\mu\to e+\gamma$, $\tau\to e + \gamma$ and $\tau \to \mu +
\gamma$ have been calculated both in the cases of $\mathbf{R} =
\mathbf{1}$ and of complex matrix $\mathbf{R}$.

Our results show that if $\mathbf{R} = \mathbf{1}$, taking into
account the effects of the RG running of the neutrino mixing
parameters in the case of relatively large $\tan\beta$ and/or $m_1$,
can change significantly the predictions for the $l_i\to l_j + \gamma$
decay branching ratios $\text{BR}(l_i\to l_j + \gamma)$.  The
magnitude of this change depends strongly on the value of the Majorana
CP-violation phase $\alpha$.  More specifically, we find that for $
0.05 \ltap m_1 \ltap 0.30$ eV and $30 \ltap \tan\beta \ltap 50$, the
predicted branching ratios $\text{BR}(\mu \to e + \gamma)$ and
$\text{BR}(\tau \to e + \gamma)$, can be enhanced by the effects of
the RG running of $\theta_{ij}$ and $(m_{2(3)} - m_1)$ by 1 to 3
orders of magnitude if $\pi/4 \ltap \alpha \ltap \pi$, while
$\text{BR}(\tau \to \mu + \gamma)$ can be enhanced by up to a factor
of 10.  For $\alpha \cong 0$, the combined effect of the RG running of
$\theta_{12}$, $\theta_{23}$ $\theta_{13}$ and $m_j$ on
$\text{BR}(l_i\to l_j + \gamma)$ was found to be relatively small
(essentially negligible) even for $ m_1 \sim (0.2 - 0.3)$ eV and
$\tan\beta \sim 50$. 

We have considered further the case of non-trivial complex matrix
$\mathbf{R}$  having the form given by eq. (\ref{exp}).  In this case
$\mathbf{R}$ can be parametrised by three real parameters $a,b,c$. The
form of $\mathbf{R}$ we have employed is the most general relevant one
for the study of the LFV decays $l_i \to l_j + \gamma$ in the case of
QD heavy RH neutrinos.  As it has been shown in~\cite{PPY03},
successful leptogenesis requires $abc \neq 0$ and $10^{-5}\ltap |abc|
\ll 1$.  We find that in the case of $\tan\beta \sim 10$ and $m_1 \sim
0.06$ eV, $\ie$, when the RG effects are negligible, the branching
ratios $\text{BR}(\mu\to e+\gamma)$ and $\text{BR}(\tau\to e+\gamma)$
can be significantly enhanced for any value of the Majorana phase
$\alpha$ by the effect of complex $\mathbf{R}$~\cite{PPY03}, provided
$r\equiv \sqrt{a^2+b^2+c^2} \gtap 5\times 10^{-3}$ (Fig.
\ref{m1-lfv-3}).  The corresponding enhancement factor can be as large
as $\sim 100$ for $r \cong 5\times 10^{-2}$.  For relatively large
values of $\alpha$, $\eg$ $\alpha \sim [\pi/2,\pi]$, however, and
depending on the signs and values of the parameters $a,b,c$,
$\text{BR}(\mu\to e+\gamma)$ and $\text{BR}(\tau\to e+\gamma)$ can
also be suppressed as a consequence of a partial cancellation between
the different contributions in the amplitudes of these two processes.
The effect of complex $\mathbf{R}$ on $\text{BR}(\tau\to \mu+\gamma)$
is minor for $r \ltap 5\times 10^{-2}$ and $m_1 < 0.10$ eV.  In the
case of $a,b,c$ satisfying $|a|,|b|,|c| \leq 0.10$, the contributions
due to $\mathbf{R}$ in $\text{BR}(\mu\to e+\gamma)$ and
$\text{BR}(\tau\to e+\gamma)$ become dominant at $m_1 \gtap 0.01$ eV;
for $\text{BR}(\tau\to \mu +\gamma)$ they become dominant at $m_1
\gtap 0.10$ eV. Similar results are valid in the case of $\tan\beta
\sim 50$.  For $\pi/2 \ltap \alpha \leq \pi$ in this case, the
enhancement of $\text{BR}(\mu\to e+\gamma)$ and $\text{BR}(\tau\to
e+\gamma)$ at $m_1 \gtap 0.02$ eV due to the RG effects is so strong
that the contributions of the complex $\mathbf{R}$ can lead to further
increase of $\text{BR}(\mu\to e+\gamma)$ and $\text{BR}(\tau\to
e+\gamma)$ by not more than approximately an order of magnitude. For
sufficiently large $\alpha$, $\eg$ $\alpha \sim [\pi/2,\pi]$, these
contributions can also compensate partially the RG effect of
enhancement leading to significantly smaller $\text{BR}(\mu\to
e+\gamma)$ and $\text{BR}(\tau\to e+\gamma)$ than in the case of
$\mathbf{R} = \mathbf{1}$ (Fig.~\ref{m1-lfv-2}).  Qualitatively
similar results are valid also for $\text{BR}(\tau\to \mu +\gamma)$.
If the Majorana phase $\beta_M$ is sufficiently large, it can have an
effect on the contributions to $\text{BR}(l_i \to l_j + \gamma)$ due
to the complex matrix $\mathbf{R}$, similar to that of the phase
$\alpha$ discussed above.

The results of the present study show that taking into account the
effects of the Majorana CP-violation phases in the PMNS matrix and of
the RG evolution of the light neutrino mixing parameters -- masses and
PMNS mixing angles, can have a remarkably large impact on the
predicted rates of the LFV decays $\mu \to e + \gamma$ and $\tau \to e
+ \gamma$, $\etc$. in MSSM with right-handed neutrinos and see-saw
mechanism of neutrino mass generation.  The impact of these effects on
the $\tau \to \mu + \gamma$ decay rate is typically smaller and
possible only for $m_1 \gtap 0.10$ eV.

\vspace{1cm}
{\bf Acknowledgements.} 
S.T.P. would like to thank the Institute of
Nuclear Theory at the University of Washington, 
Seattle, for its hospitality, and the 
U.S. Department of Energy for partial support, 
during the completion of the present study.
This work was supported in part by the Italian 
INFN  under the program ``Fisica Astroparticellare'' 
(S.T.P. and T.S.).

%
\section*{\large{APPENDIX A}}
\label{sol-rge}
%
%
\hskip 1.0truecm
In the present Appendix we derive approximate 
analytic expressions for some of the
neutrino mixing parameters at the scale $M_R$.
The renormalisation group (RG) 
equation for the neutrino mass matrix $m_{\nu}$
in MSSMRN has the form (see, $\eg$,~\cite{RGrunU})
\begin{align}
\frac{d}{d\ln \mu_R}m_{\nu}=
\frac{1}{8\pi^2}\left[
\left\{-4\pi\left(3\alpha_2+\frac{3}{5}\alpha_1\right)
+\mathrm{tr}(3Y_U^{\dagger}Y_U)\right\}m_{\nu}
+\frac{1}{2}\left\{
(Y_E^{\dagger}Y_E)^Tm_{\nu}+m_{\nu}(Y_E^{\dagger}Y_E)
\right\}
\right]\;,
\end{align}
%
where $\mu_R$ is the RG energy scale, 
$\alpha_j = g_j^2/(4\pi)$, where $g_3$, $g_2$ and $g_1$ 
are the $SU(3)^c$, $SU(2)_{L}$ and $U(1)_Y$ gauge coupling 
constants, respectively, and
$Y_U$ is the matrix of Yukawa couplings of up-quarks.
This equation admits the approximate 
solution~\cite{LolaE99}
\begin{align}
m_{\nu}(M_R)\simeq J(M_R)
\begin{pmatrix}
1&&\\
&1&\\
&&1+\epsilon_{\tau}
\end{pmatrix}
m_{\nu}(M_Z)
\begin{pmatrix}
1&&\\
&1&\\
&&1+\epsilon_{\tau}
\end{pmatrix}\;,
\label{mMR}
\end{align}
%
where $J(M_R)$ and $\epsilon_{\tau}$ are given by
\begin{align}
\label{JMR}
J(M_R)=&\exp\left[
\frac{1}{8\pi^2}\int_{\ln M_Z}^{\ln M_R}
\left(\mathrm{tr}(3Y_U^{\dagger}(\mu)Y_U(\mu))
-4\pi\left(3\alpha_2(\mu)+\frac{3}{5}\alpha_1(\mu)\right)
\right)d(\ln \mu)\right]\;,\\
\epsilon_{\tau}=&
1-\left(\frac{M_Z}{M_R}\right)^{(1/8\pi^2)(1+\tan^2\beta)(m_{\tau}/v)^2}\;,
\label{etau}
\end{align}
%
$m_{\tau}$ being the $\tau^{\pm}$ mass. 
Note that $\epsilon_{\tau}$ takes positive values and is
a monotonically increasing function of $\tan\beta$.
The PMNS matrix at $M_R$, $U(M_R)$,
is obtained by diagonalising $m_{\nu}(M_R)$:
\begin{align}
U(M_R)^Tm_{\nu}(M_R)U(M_R)
=\mathrm{diag}(m_1(M_R),m_2(M_R),m_3(M_R))\;.
\label{UMR}
\end{align}
%
Using eqs. (\ref{eq:Upara}) and 
(\ref{mMR}) - (\ref{UMR}) it is possible 
to derive the following approximate expressions for
the solar neutrino mixing angle $\theta_{12}$ and
$\Delta m^2_{21}$ at $M_R$:
\begin{align}
\tan\theta_{12}(M_R) \simeq
\frac{\left|\sin\theta_{12}(M_Z)\cos\theta_x
+\cos\theta_{12}(M_Z)\sin\theta_x e^{-i\alpha/2}\right|}
{\left|\cos\theta_{12}(M_Z)\cos\theta_x
-\sin\theta_{12}(M_Z)\sin\theta_x e^{-i\alpha/2}\right|}\;,
\label{solMR-solmz}
\end{align}
%
and 
\begin{align}
&\Delta m^2_{21}(M_R) \simeq J^2(M_R)
\frac{\Delta m_{21}^2(M_Z)
+4\epsilon_{\tau}m^2_1(M_Z)\sin^2\theta_{23}(M_Z)
\cos2\theta_{12}(M_Z)}
{\cos 2\theta_x}\;,
\label{eq_massdef_MR}
\end{align}
%
where the angle $\theta_x$ is determined by the equation 
\begin{align}
\label{thx}
\tan 2\theta_x \simeq
-~\frac{4\epsilon_{\tau}m_1^2(M_Z)\sin^2\theta_{23}(M_Z)
\sin2\theta_{12}(M_Z)\cos\frac{\alpha(M_Z)}{2}}
{\Delta m^2_{21}(M_Z)+ 4\epsilon_{\tau}m^2_1(M_Z)\sin^2\theta_{23}(M_Z)
\cos2\theta_{12}(M_Z)}\;.
\end{align}
%
As can be shown, eqs. (\ref{eq_massdef_MR}) and
(\ref{thx}) imply
$\cos 2\theta_x>0$, $\sin 2\theta_x <0$, provided
we keep the ordering $m_1(M_R) < m_2(M_R)$ 
of the masses $m_1$ and $m_2$ at $M_R$.
It follows, in particular, from eq. (\ref{solMR-solmz}) that
\begin{align}
\sin^22\theta_{12}(M_R)\leq \sin^22\theta_{12}(M_Z)\;.
\end{align}
%
For $\alpha(M_Z)=0$ and $\alpha(M_Z) = \pi$, for instance,
we have
\begin{equation}
\sin^22\theta_{12}(M_R)=
\begin{cases}
\sin^22(\theta_{12}(M_Z)+\theta_x)\;,& \alpha=0\;,\\
\sin^22\theta_{12}(M_Z)\;, &\alpha=\pi\;.
\end{cases}
\end{equation}
%
\indent The mixing angle $\theta_{13}$ and the Dirac CP-violation phase
$\delta$ are also affected by the RG evolution from
$M_Z$ to $M_R$.  If, $\eg$ we set $s_{13}=0$ at $M_Z$,
we get for $s_{13}$ and $\delta$ at $M_R$: 
\begin{align}
s_{13}(M_R)
\simeq \frac{J^2(M_R)\epsilon_{\tau} 
m_1(M_Z)m_3(M_Z)\sin 2\theta_{12}(M_Z)\sin 2\theta_{23}(M_Z)}
{m_3^2(M_R)-m_1^2(M_R)}
\sin\frac{\alpha(M_Z)}{2}\;,
\end{align}
%
and 
\begin{align}
\delta(M_R) \simeq 
\xi_1+\xi_2-\frac{\pi}{2}+\frac{\alpha(M_Z)}{2}-\beta_M(M_Z)\;,
~~~{\rm for}~\alpha(M_Z)\neq 0\;,
\end{align}
%
where
\begin{align}
&\xi_1 = \arg(c_{12}(M_Z)c_x - s_{12}(M_Z)s_xe^{-i\alpha(M_Z)/2})\;,
\nonumber\\
&\xi_2 = \arg(s_{12}(M_Z)c_x + c_{12}(M_Z)s_xe^{i\alpha(M_Z)/2})\;.
\end{align}
%
The angle $\theta_{13}(M_R)$ takes its largest value for
$\alpha(M_Z)=\pi$
%
and in this case $\delta(M_R)\cong - \beta_M(M_Z)$.
In contrast, if $\alpha=0$, the correction 
to $\sin\theta_{13}$ due to the RG running 
is negligible.

%

%
\end{document}